\RequirePackage{fix-cm}
\documentclass[twocolumn,epjc3]{svjour3}  
\smartqed  
\RequirePackage{graphicx}
\journalname{Eur. Phys. J. C}
\usepackage{lipsum}
\usepackage{mathtools}
\usepackage{cuted}
\usepackage[dvipsnames]{xcolor}
\usepackage{booktabs}
\usepackage{xurl}
\usepackage[
  colorlinks=true,
  linkcolor=blue,
  citecolor=blue,
  urlcolor=blue,
  breaklinks=true
]{hyperref}
\usepackage{subcaption}
\numberwithin{equation}{section}
\usepackage[T1]{fontenc}
\usepackage[utf8]{inputenc} 
\usepackage{lmodern}
\usepackage{microtype}
\usepackage[english]{babel}
\begin{document}

\title{Entropic Chaplygin-Gas cosmology and late-universe tension diagnostics}



\author{Kelvis A. Kulhkamp\textsuperscript{1}
        \and
        Carlos H. Coimbra-Ara\'ujo\textsuperscript{1,2} 
        \and 
        Abraão J. S. Capistrano\textsuperscript{1,2}
        \and
        Luiz A. Cabral\textsuperscript{2,3}
        \and
        Jos\'e A. P. F. Mar\~ao\textsuperscript{4}
}



\institute{\textsuperscript{1}Applied physics graduation program, UNILA, 85867-670, Foz do Iguaçu-PR, Brazil \label{addr1}
            \\
            \email{kelvisandrei@gmail.com}
           \\[4pt]
            \textsuperscript{2}
            Universidade Federal do Paraná, 85950-000, Palotina-PR, Brazil \label{addr2}
            \\
            \email{carlos.coimbra@ufpr.br} \\
            \email{capistrano@ufpr.br}
            \\[4pt]
            \textsuperscript{3}
            Curso de F\'isica, Setor Cimba, Universidade Federal do Tocantins, 77824-838, Aragua\'ina-TO, Brazil\label{addr3}
            \\
            \email{cabral@uft.edu.br}
            \\[4pt]
            \textsuperscript{4}
            Centro Tecnol\'ogico, Departamento de Matem\'atica,
            Universidade Federal do Maranh\~ao, 65085580, S\~ao Lu\'is-MA, Brazil\label{addr4}
            \\
            \email{jose.marao@ufma.br} 
           }

\date{Received: date / Accepted: date}

\maketitle

\begin{abstract}
\begingroup
\sloppy
\emergencystretch=2em
We investigate an entropic generalized Chaplygin-gas (gCg) cosmology in which a single dark fluid accounts for both dark matter and dark energy. We constrain the model using DESI baryon acoustic oscillations, redshift-space-distortion measurements, compressed Planck 2018 distance and lensing information, and Type Ia supernova likelihoods with PantheonPlus+SH0ES (PPS) and DES--Dovekie. For Planck+DESI+RSD+PPS, the entropic gCg model yields a Hubble constant of $H_{0}=68.64\pm0.52\,{\rm km\,s^{-1}\,Mpc^{-1}}$ and the amplitude of matter fluctuations parameter of $S_{8}=0.797\pm0.021$, compared with $H_{0}=67.31\pm0.28\,{\rm km\,s^{-1}\,Mpc^{-1}}$ and $S_{8}=0.809\pm0.021$ in $\Lambda$CDM. The simultaneous upward shift in $H_{0}$ and downward shift in $S_{8}$ occurs in the directions required for a partial alleviation of both tensions. We also show that when PPS is replaced by DES--Dovekie, the parameter shifts are smaller, the best-fit improvement is negligible, and the Bayesian evidence favours $\Lambda$CDM. Thus, the entropic gCg model provides a potentially interesting simultaneous shift of the Hubble constant and the matter-clustering amplitude, but its statistical support depends on the adopted supernova likelihood.
\par
\endgroup

\end{abstract}

\section{Introduction}
\label{sec:introduction}

In the late 1990, robust pioneering observations at high redshifts of Type Ia supernovae \cite{DE1,RIESS,RIESS2,tonry} and from the Cosmic Microwave Background~(CMB) \cite{melchiorri,lange,jaffe,halverson,netterfield} showed that the universe is expanding rapidly. To explain this phenomenon, cosmologists have proposed the existence of an energetic component with negative pressure in the universe known as dark energy~(DE). For more than twenty years till now $\Lambda$CDM cosmological paradigm has been both the simplest and the most eminent and effective explanation to get along with the observed accelerated expansion of the universe \cite{jaffe,sahnistaro,percival,alamet,kowalski,izzo,efstathiou,allen,baxter,chavez,aghanim}. Even with this success, $\Lambda$CDM model has important detriments that must be taken into consideration. One of them is that the $\Lambda$CDM model lacks a fundamental theoretical explanation about the nature of the cosmological constant $\Lambda$~\cite{Weinberg1989CosmologicalConstant}. Another issue is that $\Lambda$CDM model fail to give any hint on the nature of the cold Dark Matter~(CDM) \cite{nemi,santos,kumar,velten,sultana,siva,noza}. Moreover, given the lack of a definitive understanding of DE nature, numerous models have been proposed in an effort to offer alternative explanations to this unresolved problem. Some examples are the models with cosmological constant \cite{carroll}, homogeneous scalar field $\Phi$ with an effective potential $v(\Phi)$ \cite{caldwell,saini}, models of matter X characterized by a equation of state $p=w\rho$, with $-1\leq w <0$ \cite{peebles2}, effects of extra-dimensions and modified gravity \cite{sahni,Capistrano2021PRD,Capistrano2021PDU2,Capistrano2021PDU,Capistrano2021CQG,Capistrano2022Galaxies,Capistrano2023Universe,Andrade2024MNRAS,Capistrano2024PRD,Capistrano2026PRD}, or exotic fluids like the Chaplygin gas \cite{karman}. A review of equivalent descriptions of DE cosmologies and cosmographic tests can be found in \cite{Bamba2012DarkEnergy}. 

While the cosmological constant remains the simplest and most widely adopted model for dark energy~(DE), a variety of dynamical models have been proposed to address theoretical and observational challenges. Recent DESI-era reviews emphasize both the emerging evidence for departures from $\Lambda$CDM and the sensitivity of those indications to dataset choice and modeling assumptions \cite{Li:2026asg}. Among these alternatives, the Chaplygin gas and its generalizations have attracted attention because they interpolate between dust-like matter at early times and a DE-dominated regime at late times \cite{KAMENSHCHIK2001265,Bilic2002,Bento2002}. Chaplygin-gas models may provide an observational framework for accelerated expansion and the dark-matter problem \cite{Bento2003,Aurich2018,Li2018}, and offer new routes for addressing the cosmological constant problem \cite{Bertolami2023}, the Hubble tension \cite{Yang2019}, the Swampland conjectures \cite{Bertolami2024}, the accretion of cosmic dark fluids \cite{Reis2026}, and inflation \cite{Bertolami2006}.

Despite these attractive features, unified Chap-lygin-gas cosmologies face important theoretical and observational difficulties, particularly at the perturbative level. In their simplest adiabatic realization, generalized Chaplygin-gas models may generate large effective sound speeds that induce oscillations and instabilities in the matter power spectrum, suppress structure formation, and lead to tensions with CMB and large-scale-structure observations \cite{Sandvik2004,Bean2003,Amendola2003}. Consequently, viable scenarios generally require tightly constrained parameter regions or non-adiabatic/entropic extensions capable of reducing the effective clustering sound speed while preserving the successful background evolution. In general, interacting or unified descriptions of the dark sector remain actively investigated, with recent DESI-era work stressing that apparent preferences must be tested for robustness across likelihood choices \cite{kumar2017,valentino2017,yang2017,pan2018,yang2018,yang2019_2,pan2019,Li:2026xaz}. Chaplygin-gas models based on this idea are known as unified dark matter--dark energy (UDM) models \cite{Bilic2002,Bento2002,Mamon2022EPJC,Goray2025EPJC}.

In the present paper we aim at investigating the entropic UDM generalized Chaplygin gas model. In this entropic version, the pressure is not solely a function of density but also depends on entropy, allowing for a more flexible and thermodynamically consistent description of cosmic evolution. This model has been explored in various contexts, including scalar field representations and holographic scenarios, offering intriguing possibilities for explaining the accelerated expansion of the universe and the behavior of cosmological perturbations (see \cite{kumar_dubey,Bertolami2023,paul} and the references therein). The layout of the paper is as follows: in Section \ref{sec: UDM} we present the cosmological background dynamics associated with generalized and entropic Chaplygin-gas cosmologies, emphasizing the unified dark-sector interpretation and the corresponding expansion history. We also discuss the evolution of the equation-of-state~(EoS) parameter, the Hubble expansion rate, and the deceleration parameter in comparison with the standard $\Lambda$CDM scenario. In Section \ref{sec:perturbations} we investigate the general perturbative regime and derive the linear Einstein equations relevant for scalar cosmological perturbations in the presence of an exotic unified dark fluid. In Section \ref{sec:data_likelihoods} we describe the observational datasets used and overall numerical implementation adopted in the statistical analysis. In Section \ref{sec:tension_diag} we discuss the observational constraints and tension diagnostics obtained from the combined analyses involving Planck Cosmic Microwave Background~(CMB) measurements \cite{Planck2020Parameters}, baryon acoustic oscillation~(BAO) observations from DESI and related surveys \cite{DESI2024BAO,eBOSS2021}, Type Ia supernova~(SNIa) compilations such as PantheonPlus+SH0ES~(PPS)\cite{Riess2022SH0ES} and DES--Dovekie \cite{Popovic2025Dovekie}, redshift-space distortion~(RSD) growth measurements \cite{Beutler2012RSD,Howlett2015RSD}, and CMB lensing information from the Planck collaboration \cite{ChenHuangWangPlanckDistancePriors}. 

The Bayesian parameter estimation is performed using the \texttt{MultiNest} nested-sampling algorithm \cite{Feroz2009MultiNest,Feroz2019} and the \texttt{Tensiometer} code \cite{Raveri2020Tensiometer} for model tension diagnosis. In addition, we examine the relative statistical performance of the cosmological models using complementary information-theory estimators as the Akaike Information Criterion~(AIC) \cite{akaike1974}, the Bayesian Information Criterion~(BIC) \cite{schwarz1978}, and the Deviance Information Criterion~(DIC) \cite{spiegelhalter2002}. Finally, conclusions and propescts are made in the conclusion section. Hereon we will use natural units, $c=G=1$, and our spacetime has $(-+++)$ signature. Greek indices run from 0 to 3.

\section{Background UDM Chaplygin cosmology}\label{sec: UDM}

The exotic pure Chaplygin gas (Cg) has an original EoS of the type
\begin{equation}
    p_{Cg}=-\frac{A}{\rho}\;,
\end{equation}
where $p_{Cg}$ is the gas pressure, $\rho$ is the energy density, with $\rho>0$, and $A$ is a positive constant. It is well known that pure Chaplygin gas is able to explain the accelerated expansion of the universe, but fails to explain the formation of structures. Given this scenario, \cite{Bilic2002} and \cite{Bento2002} proposed a modification of EoS, which took the form
\begin{equation}
\label{gCg}
    p_{gCg}=-\frac{A}{\rho^\alpha}\;,
\end{equation}
with $0\leq\alpha\leq1$ so that the adiabatic sound speed remains subluminal. If $\alpha=1$ we recover the pure Chaplygin gas. For $\alpha=0$, the density becomes
\begin{equation}
\rho_{\rm gCg}(a)=A+Ba^{-3},
\label{eq:lcdm_limit_gcg}
\end{equation}
which is exactly equivalent at the background level to a cosmological constant plus pressureless matter; no special choice $A=1$ is required. Therefore, this new EoS is known as generalized Chaplygin gas (gCg) and is a phenomenological extension of the pure Cg.

The background equations are obtained using the standard Friedman-Robertson-Walker (FLRW) metric, where the line element with comoving coordinates is
\begin{equation}
\label{FLRW}
ds^2=-dt^2+a^2\left[\frac{dr^2}{1-kr^2}+r^2(d\theta^2+\sin^2\theta d\phi^2)\right],
\end{equation}
where $a=a(t)$ is the scale factor, $(t,r,\theta,\phi)$ is the comoving system of coordinates and $k$ is the spatial section curvature parameter. We begin from the continuity equation

\begin{equation}
\label{consME1}
   \nabla_\nu T^{\mu \nu}=\dot{\rho}+3\frac{\dot{a}}{a}(p+\rho)=0\;,
\end{equation}
where the dot refers to the time derivative and $\nabla_\nu$ is the usual operator for a covariant derivative. Here, $T_{\mu \nu}=(\rho + p)u_{\mu}u_\nu+pg_{\mu\nu}$ is the stress tensor of a perfect fluid, with $u_\mu$ being the fluid velocity four-vector. Considering the EoS in Eq.\eqref{gCg}, we obtain

\begin{equation}
    d\rho=-3\frac{da}{a}\left(\frac{\rho^{\alpha+1}-A}{\rho^\alpha}\right).
   \end{equation}
When integrated this equation results in an expression for the density

\begin{equation}
\label{densidad.gCg}
    \rho_{gCg}=\left(A+\frac{B}{a^{3(1+\alpha)}}\right)^{\frac{1}{1+\alpha}},
\end{equation}
\noindent where $B$ is an integration constant. The evolution of the dark fluid parameter $w$ of the gCg EoS with $p=w\rho$ is now given by

\begin{equation}
\label{Eos gCg}
    w_{gCg}=\frac{p_{gCg}}{\rho_{gCg}}=-\left[1+\frac{B}{A}a^{-3(1+\alpha)}\right]^{-1}.
\end{equation}

\noindent Analyzing the density of Eq.\eqref{densidad.gCg} and considering $\alpha=1$, the density expression can be written as $\rho=\sqrt{A+\frac{B}{a^6}}$ and is related to the density evolution for a pure Cg. Note that for a universe in its initial stage of evolution $a \rightarrow 0$ and therefore $\frac{B}{a^6}\gg A$, we obtain that the density is expressed by $\rho\approx\frac {\sqrt{B}}{a^3}$, corresponding to the evolution of matter. On the other hand, if we consider $a \rightarrow \infty$ we get the relation $A \gg \frac{B}{a^6}$, obtaining that the density $\rho \approx \sqrt{A}$ is constant, which could be related to the density of the late effect of DE in current time. This analysis leads to the conceptual conclusion that there would be a unification of the dark sector in a single EoS.

A currently important cosmological background parameter is the Hubble parameter $H=H(t)$, which is interpreted as the relative rate of expansion of the universe at a given time. Since from Eq.\eqref{consME1} the density $\rho$ can assume different evolution patterns, e.g., a universe dominated by radiation or by barions (i.e., respectively, $\rho_r=\rho^0_r a^{-4}$, $\rho_b=\rho^0_b a^{-3}$), or even by the dark sector then, for a flat universe where $k=0$, and assuming that the gCg represents the dark sector (i.e., $\rho_{de+dm}=\rho_{gCg}$), we can write the Friedmann equation as 

\begin{equation}
H^2=\frac{8\pi }{3}\left[\left(A+Ba^{-3(1+\alpha)}\right)^{\frac{1}{1+\alpha}}+\rho^0_b a^{-3}+\rho^0_r a^{-4}\right].
\end{equation}

\noindent And since the critical density is defined by $\rho_{crit}=\frac{3H_0^2}{8\pi }$ and the density parameter by $\Omega_i=\frac{\rho^0_i}{\rho_{crit}}$, the above equation becomes
\begin{equation}
\label{hubble.param.udm1}
    H^2=H_0^2\left[\frac{\left(A+Ba^{-3(1+\alpha)}\right)^{\frac{1}{1+\alpha}}}{\rho_{crit}}+\Omega^0_b a^{-3}+\Omega^0_r a^{-4}\right],
\end{equation}

\noindent where $H_0$ is the current value of the Hubble parameter. In addition, we use Eq.\eqref{densidad.gCg} to obtain

\begin{equation}
     \rho^0_{de+dm}=\rho(1)=(A+B)^{\frac{1}{1+\alpha}},
\end{equation}

\noindent and since $\rho_{crit}=\rho^0_{de+dm}/\Omega^0_{de+dm}$, Eq.\eqref{hubble.param.udm1} becomes

\begin{eqnarray}
\label{hubble.param.udm2}
 H^2&=&H_0^2\Omega_{de+dm}^0\left(\frac{1+\tilde{A}a^{-3(\alpha+1)}}{1+\tilde{A}}\right)^{\frac{1}{1+\alpha}}\nonumber \\
 &&+H_0^2(\Omega^0_ba^{-3}+\Omega^0_r a^{-4}), 
\end{eqnarray}

\noindent where we made $\tilde{A}=\frac{B}{A}$ so that the free parameters reduce from three to two ($\tilde{A}$ and $\alpha$).

\begin{figure}
    \centering
    \includegraphics[width=1\linewidth]{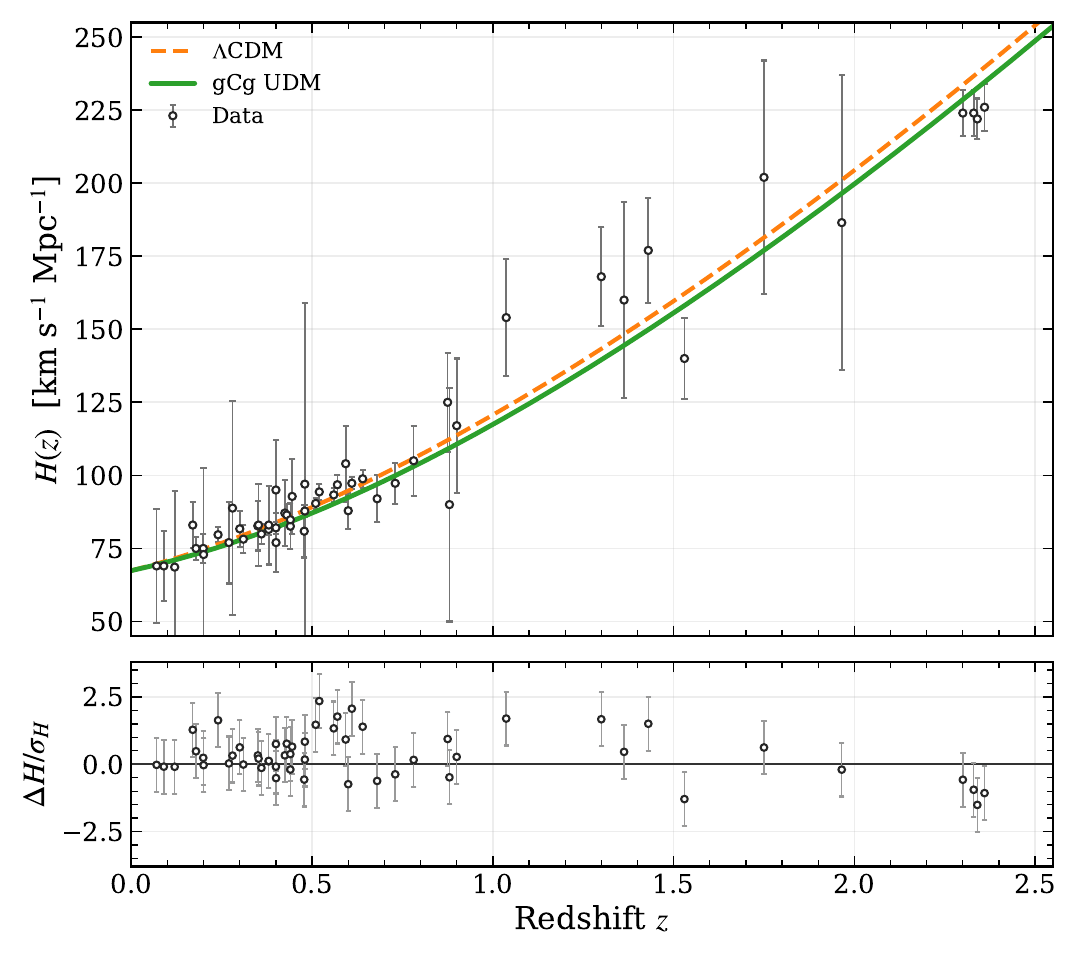}
    \caption{ In the Top panel, we show the evolution of the Hubble parameter $H(z)$ as a function of redshift $z$ for the $\Lambda$CDM model (orange dashed line) and gCg UDM model (green solid line), compared with observational $H(z)$ data (black points with $1\sigma$ error bars). The gCg model is computed using the baseline parameter values $\tilde{A}=0.3$ and $\alpha=0.12$, while the $\Lambda$CDM baseline adopts $\Omega_{m0}=0.315$ and $H_0=67.4\,\mathrm{km\,s^{-1}\,Mpc^{-1}}$. In the Bottom panel, it shows normalized residuals $(H_{\mathrm{model}} - H_{\mathrm{data}})/\sigma_H$ for the same datasets, illustrating the statistical consistency of both models with observational constraints.}
    \label{fig:Hz_with_residuals}
\end{figure}

\begin{table}
\centering
\renewcommand{\arraystretch}{1}
\begin{tabular}{cccccccc}
\hline
$z$ & $H(z)$ & $\sigma_H$ & Ref. & $z$ & $H(z)$ & $\sigma_H$ & Ref. \\ 
\hline
0.07    & 69.0  & 19.6 & \cite{graficohubble4} & 0.48  & 97.0   & 62.0 & \cite{graficohubble2} \\
0.09    & 69.0  & 12.0 & \cite{graficohubble1} & 0.51  & 90.4   & 1.9  & \cite{alamet} \\    
0.12    & 68.6  & 26.2 & \cite{graficohubble4} & 0.52  & 94.35  & 2.65 & \cite{graficohubble10} \\
0.17    & 83.0  & 8.0  & \cite{graficohubble1} & 0.56  & 93.33  & 2.32 & \cite{graficohubble10} \\
0.179   & 75.0  & 4.0  & \cite{graficohubble3} & 0.57  & 96.8   & 3.4  & \cite{graficohubble14} \\
0.199   & 75.0  & 5.0  & \cite{graficohubble3} & 0.593 & 104.0  & 13.0 & \cite{graficohubble3} \\
0.2     & 72.9  & 29.6 & \cite{graficohubble4} & 0.60  & 87.9   & 6.1  & \cite{graficohubble13} \\
0.24    & 79.69 & 2.65 & \cite{graficohubble8} & 0.61  & 97.3   & 2.1  & \cite{alamet} \\
0.27    & 77.0  & 14.0 & \cite{graficohubble1} & 0.64  & 98.82  & 2.99 & \cite{graficohubble10} \\
0.28    & 88.8  & 36.6 & \cite{graficohubble4} & 0.68  & 92.0   & 8.0  & \cite{graficohubble3} \\
0.3     & 81.7  & 6.22 & \cite{graficohubble9} & 0.73  & 97.3   & 7.0  & \cite{graficohubble13} \\
0.31    & 78.17 & 4.74 & \cite{graficohubble10} & 0.781 & 105.0  & 12.0 & \cite{graficohubble3} \\
0.35    & 82.7  & 8.4  & \cite{graficohubble11} & 0.875 & 125.0  & 17.0 & \cite{graficohubble3} \\
0.352   & 83.0  & 14.0 & \cite{graficohubble3} & 0.88  & 90.0   & 40.0 & \cite{graficohubble2} \\
0.36    & 79.93 & 3.39 & \cite{graficohubble10} & 0.9   & 117.0  & 23.0 & \cite{graficohubble1} \\
0.38    & 81.5  & 1.9  & \cite{alamet} & 1.037 & 154.0  & 20.0 & \cite{graficohubble3} \\
0.3802  & 83.0  & 13.5 & \cite{graficohubble6} & 1.3   & 168.0  & 17.0 & \cite{graficohubble1} \\
0.40    & 82.04 & 2.03 & \cite{graficohubble10} & 1.363 & 160.0  & 33.6 & \cite{graficohubble5} \\
0.4     & 95.0  & 17.0 & \cite{graficohubble1} & 1.43  & 177.0  & 18.0 & \cite{graficohubble1} \\
0.4004  & 77.0  & 10.2 & \cite{graficohubble6} & 1.53  & 140.0  & 14.0 & \cite{graficohubble1} \\
0.4247  & 87.1  & 11.2 & \cite{graficohubble6} & 1.75  & 202.0  & 40.0 & \cite{graficohubble1} \\
0.43    & 86.45 & 3.68 & \cite{graficohubble8} & 1.965 & 186.5  & 50.4 & \cite{graficohubble5} \\
0.44    & 82.6  & 7.8  & \cite{graficohubble13} & 2.3   & 224    & 8    & \cite{graficohubble18} \\    
0.44    & 84.81 & 1.83 & \cite{graficohubble10} & 2.33  & 224    & 8    & \cite{graficohubble15} \\
0.44497 & 92.8  & 12.9 & \cite{graficohubble6} & 2.34  & 222    & 7    & \cite{graficohubble16} \\
0.4783  & 80.9  & 9.0  & \cite{graficohubble6} & 2.36  & 226    & 8    & \cite{graficohubble17} \\
0.48    & 87.79 & 2.03 & \cite{graficohubble10} &       &        &      &                        \\
\hline
\end{tabular}
\caption{Observational data for the Hubble parameter for different values of redshift. Table partially based on \cite{graficohubble6}.}
\label{tab:table1}
\end{table}

In Fig.\eqref{fig:Hz_with_residuals}, the upper panel shows the redshift evolution of the Hubble parameter $H(z)$. The orange dashed curve represents the reference $\Lambda$CDM cosmology, while the green solid curve represents gCg UDM model. The black circular points with vertical error bars correspond to the observational $H(z)$ measurements, with the error bars denoting the reported $1\sigma$ uncertainties. The two theoretical curves are close over the full plotted redshift range, indicating that the gCg model reproduces the background expansion history compatible with that of the standard cosmology with only mild deviations. At intermediate redshifts, the gCg prediction lies slightly below the $\Lambda$CDM curve, reflecting the different effective evolution of the unified dark fluid during the transition from matter-like to DE-like behavior.

The lower panel presents the normalized residuals, $(H_{\rm model}-H_{\rm obs})/\sigma_H$, computed with respect to the gCg model. The residuals are distributed around the horizontal zero line and most of them remain within approximately the $2\sigma$ interval. This shows that the deviations between the model prediction and the data from Table \eqref{tab:table1} are statistically compatible with the observational uncertainties, with apparently no strong redshift-dependent systematic trend.

We fixed the baseline cosmology to the values of 
$H_0=67.4\,{\rm km\,s^{-1}\,Mpc^{-1}}$, $\Omega_{b0}=0.0493$,
$\Omega_{r0}=9\times10^{-5}$, $\Omega_{k0}=0$, and
$\Omega_{m0}^{\Lambda{\rm CDM}}=0.315$, which define a Planck-like flat reference cosmology. For gCg UDM model, we fixed parameters $\tilde{A}=0.3$ and $\alpha=0.12$. These values are physically motivated because they allow the gCg density to behave approximately as pressureless matter at early times and as a negative pressure DE component at late times. With these parameters, the fit gives $\chi^2=45.42$ for $50$ degrees of freedom, corresponding to $\chi^2/{\rm dof}=0.91$. As a first test, this value indicates a statistically good background-level agreement between the gCg unified dark sector model and the observational Hubble data.

\begin{figure}[h]
    \centering
    \includegraphics[width=0.50\textwidth]{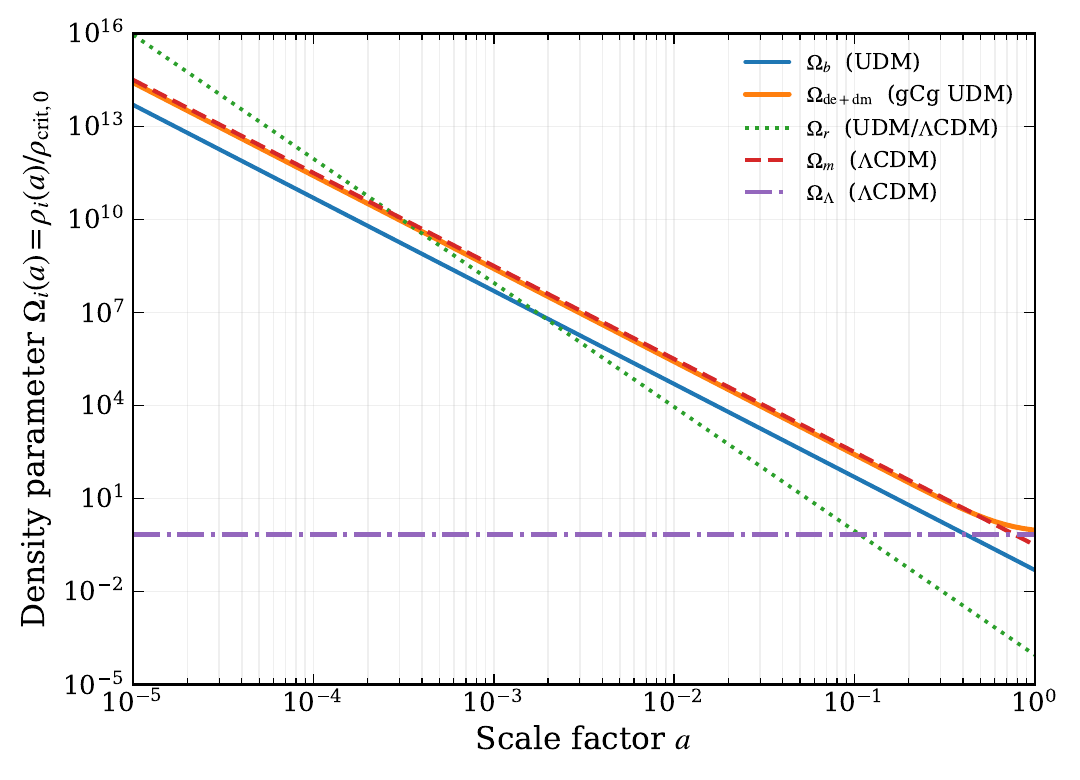}
    \caption{Evolution of the density parameters $\Omega_i(a) = \rho_i(a)/\rho_{\mathrm{crit},0}$ as a function of the scale factor $a$ in logarithmic scale. The plot compares the gCg UDM model with the standard $\Lambda$CDM scenario. The blue solid line represents baryons $\Omega_b$ (UDM), the orange solid line corresponds to the unified dark sector $\Omega_{\mathrm{de+dm}}$ (gCg UDM), the green dotted line shows radiation $\Omega_r$, the red dashed line represents matter $\Omega_m$ in the $\Lambda$CDM model, and the purple dash-dotted line corresponds to the cosmological constant $\Omega_\Lambda$. At early times, radiation dominates the energy budget, followed by a matter-dominated era. At late times, the UDM component transitions to a DE-like regime, reproducing cosmic acceleration similarly to the cosmological constant in $\Lambda$CDM.}
    \label{fig:density_parameters} 
   \end{figure}

\begin{figure}[h]
    \centering
    \includegraphics[width=0.50\textwidth]{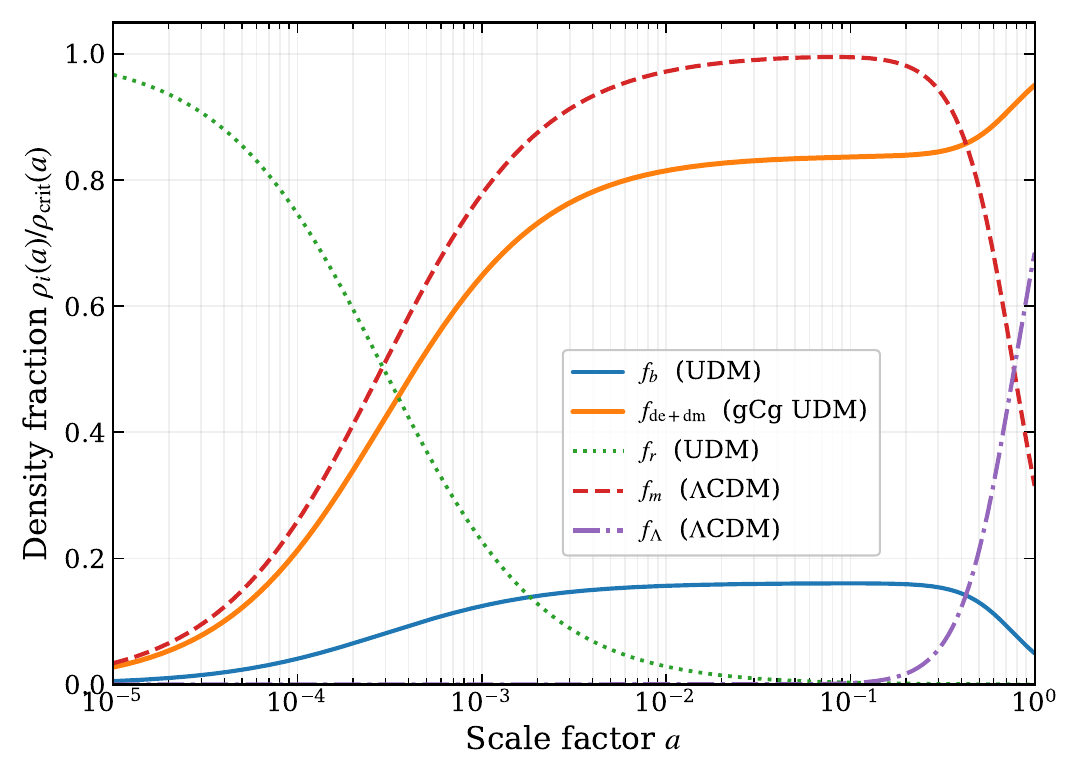}
    \caption{Evolution of the physical density fractions $f_i(a) = \rho_i(a)/\rho_{\mathrm{crit}}(a)$ as a function of the scale factor $a$. The blue solid line represents baryons $f_b$ (UDM), the orange solid line corresponds to $f_{\mathrm{de+dm}}$ (gCg UDM), and the green dotted line shows radiation $f_r$. For comparison, the red dashed line represents the matter fraction $f_m$ in the $\Lambda$CDM model, while the purple dash-dotted line corresponds to the cosmological constant $f_\Lambda$. The figure shows the transition from radiation domination to matter domination and finally to DE domination.}
    \label{fig:density_fraction} 
\end{figure}

Figs. \eqref{fig:density_parameters} and \eqref{fig:density_fraction} show the evolution of the constituent components of the universe, plotting them altogether with the evolution of the gCg density. Note that in this case, the current value of $\Omega_{de+dm}$ at $a=1$ for the gCg UDM is  $\approx 0.95$, similar to what is expected for the value of $\Omega_{de}+\Omega_{dm}$ in the $\Lambda$CDM standard model. We show the different components as distinguished by both color and line style. The blue solid line represents the baryonic component $\Omega_b$ in the UDM framework, which follows the standard matter scaling $\propto a^{-3}$ and remains subdominant at all epochs. The orange solid line corresponds to the unified dark sector $\Omega_{\mathrm{de+dm}}$ (gCg UDM). The green dotted line denotes the radiation density parameter $\Omega_r$, which dominates the energy budget at early times and decreases rapidly as $\propto a^{-4}$. The red dashed line represents the total matter component $\Omega_m$ in the $\Lambda$CDM model, which dominates during intermediate epochs and scales as $a^{-3}$. At last, the purple dash-dotted line corresponds to the cosmological constant $\Omega_\Lambda$, which remains nearly constant and dominates the late-time dynamics in the $\Lambda$CDM scenario.

In Fig.\eqref{fig:density_parameters}, we show the evolution of the density parameters $\Omega_i(a) = \rho_i(a)/\rho_{\mathrm{crit},0}$ as a function of the scale factor $a$ in logarithmic scale, comparing the gCg UDM scenario with the standard $\Lambda$CDM model. At early times (small $a \sim 10^{-5} - 10^{-3}$), the dynamics are clearly dominated by the radiation component. This is evidenced by the steepest slope among all components, consistent with the expected scaling $\Omega_r \propto a^{-4}$. In both the UDM/CDM and $\Lambda$CDM frameworks, radiation rapidly decreases as the Universe expands, becoming subdominant at later epochs. As the scale factor increases toward intermediate values ($a \sim 10^{-3} - 10^{-1}$), the matter component becomes dominant. The curves corresponding to matter, either explicitly in $\Lambda$CDM ($\Omega_m$) or effectively through the unified fluid in the UDM model, exhibit a scaling close to $a^{-3}$, reflecting non-relativistic behavior. Notably, the effective unified dark sector (gCg UDM curve) is closely to the standard matter evolution during this epoch, indicating that the UDM model successfully mimics cold dark matter at early and intermediate times.

At late times ($a \sim 10^{-1}$), the behavior of the models diverges slightly but remains qualitatively consistent. In $\Lambda$CDM case, the cosmological constant $\Omega_\Lambda$ becomes dominant, remaining nearly constant as expected for vacuum energy. On the other hand, in the UDM scenario, the unified component gradually transitions from matter-like to DE-like behavior, reproducing the late-time accelerated expansion without requiring a separate cosmological constant. This transition is smooth, and the total effective dark sector remains consistent with observational expectations.

In Fig.\eqref{fig:density_fraction}, we show the evolution of the physical density fractions $f_i(a) = \rho_i(a)/\rho_{\mathrm{crit}}(a)$ as a function of the scale factor $a$, comparing UDM scenario with the standard $\Lambda$CDM model. Unlike density parameters normalized to the present critical density, these fractions are normalized to the critical density, providing a clearer view of the relative energy budget at each epoch. For comparison purposes, the $\Lambda$CDM components are also shown. The red dashed line represents the matter fraction $f_m$, which increases from small values at early times, reaches a maximum close to unity during the matter-dominated epoch, and then decreases as DE becomes dominant. The purple dash-dotted line corresponds to the cosmological constant $f_\Lambda$, which is negligible at early times but grows rapidly at late times, eventually dominating the total energy budget.

At early times, the Universe is radiation dominated, as indicated by the green dotted curve approaching unity. Around $a \sim 10^{-4} - 10^{-3}$, a transition to matter domination occurs, where both the red dashed ($\Lambda$CDM matter) and orange solid (UDM) curves become dominant. During this epoch, the unified dark sector closely mimics the behavior of CDM. Moreover, at late times, the models exhibit their main differences while remaining phenomenologically similar. In $\Lambda$CDM, the purple dash-dotted curve ($f_\Lambda$) rapidly increases and overtakes the matter component, driving cosmic acceleration. On the other hand, in the UDM scenario, the orange solid curve smoothly evolves toward a dominant DE-like component without the need for a separate cosmological constant. 

\begin{figure}
    \centering
    \includegraphics[width=\linewidth]{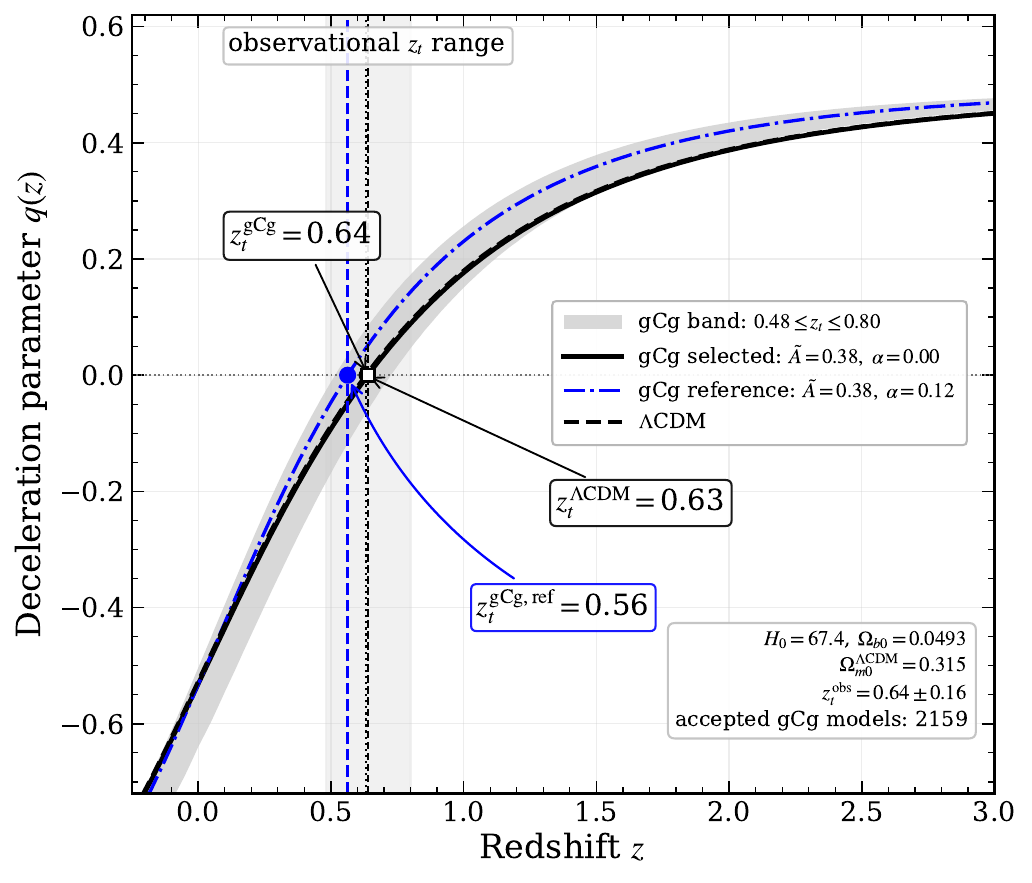}
    \caption{Evolution of the deceleration parameter $q(z)$ as a function of redshift for the $\Lambda$CDM model (black dashed curve) and the gCg model. The solid black curve represents a selected gCg realization ($\tilde{A}=0.38$, $\alpha=0.00$), while the light gray shaded region corresponds to the ensemble of viable gCg models satisfying the observational transition-redshift constraint $z_t^{\rm obs}=0.64\pm0.16$, derived from recent compilations of cosmic chronometers, BAO, and DESI $H(z)$ data. Vertical dotted lines indicate the transition redshifts defined by $q(z_t)=0$, with $z_t^{\Lambda{\rm CDM}}\simeq0.63$ and $z_t^{\rm gCg}\simeq0.64$, both lying within the observationally allowed interval $0.48\leq z_t \leq 0.80$ (shaded vertical band). For comparison purposes, the fiducial gCg model ($\tilde{A}=0.38$, $\alpha=0.12$) is shown and presents a slight deviation in the transition epoch.}
    \label{fig:deceleration}
\end{figure}

In addition, we obtain the deceleration parameter $q(a)$ in the form
\begin{equation}
\begin{aligned}
q(a)
&=
\frac{1}{2E^{2}(a)}
\Bigg\{
\Omega_{b0}a^{-3}
+2\Omega_{r0}a^{-4}
\\
&\quad
+\Omega_{{\rm gCg},0}F_{\rm gCg}(a)
\left[1+3w_{\rm gCg}(a)\right]
\Bigg\},
\end{aligned}
\label{eq:q_gcg}
\end{equation}
where $w_{gCg}$ refers to Eq.\eqref{Eos gCg} and $F_{\rm gCg}(a)$ is denoted as
\begin{equation}
F_{\rm gCg}(a)=
\left[
\frac{
1+\widetilde{A}a^{-3(1+\alpha)}
}{
1+\widetilde{A}
}
\right]^{\frac{1}{1+\alpha}}.
\label{eq:F_gcg}
\end{equation}
The squared dimensionless Hubble function $E^{2}(a)=\frac{H^2(a)}{H_0^2}$ is given by
\begin{equation}
\begin{aligned}
E^{2}(a)
&=
\Omega_{b0}a^{-3}
+\Omega_{r0}a^{-4}+\Omega_{{\rm gCg},0}F_{\rm gCg}(a),
\end{aligned}
\label{eq:E2_gcg}
\end{equation}
Thus, the deceleration parameter $q(a)$ of the UDM model with gCg presents a curve close to the $\Lambda$CDM model. The figure \eqref{fig:deceleration} shows the evolution of the deceleration parameter $q(z)$ as a function of redshift, comparing the $\Lambda$CDM model (black dashed curve) with the UDM gCg model. In addition to a fiducial gCg realization ($\tilde{A}=0.38$, $\alpha=0.12$), a band of viable gCg models is shown (gray shaded region), selected according to the observational constraint on the transition redshift $z_t=0.64\pm0.16$. The solid black curve corresponds to a representative model within this band ($\tilde{A}=0.38$, $\alpha=0.00$), which closely follows the $\Lambda$CDM behavior. The transition redshifts, defined by $q(z_t)=0$, are explicitly marked for both models, yielding $z_t^{\Lambda{\rm CDM}}\simeq0.63$ and $z_t^{\rm gCg}\simeq0.64$, in excellent agreement with the observational interval $0.48\leq z_t \leq 0.80$. The shaded vertical region highlights this allowed range, demonstrating that the selected gCg models reproduce the onset of cosmic acceleration at the correct epoch.

The curves were generated numerically from the background Friedmann equations implemented in the \texttt{Python} code, using a flat cosmology with $H_0=67.4$, $\Omega_{b0}=0.0493$, $\Omega_{r0}=9\times10^{-5}$, and $\Omega_{m0}^{\Lambda{\rm CDM}}=0.315$. For the gCg model, the density evolution $\rho_{\rm gCg}(a)=(A+Ba^{-3(1+\alpha)})^{1/(1+\alpha)}$ is used to construct $H(z)$, from which the deceleration parameter is computed via $q(z)=-1+(1+z)\,H'(z)/H(z)$. The transition redshift is determined numerically by locating the root of $q(z)$ for each parameter set. A grid over $(\tilde{A},\alpha)$ is then scanned, and only those models satisfying the observational constraint on $z_t$ are retained to build the shaded band. This procedure confirms that the gCg region is not arbitrary but statistically motivated, providing a robust comparison with $\Lambda$CDM.

From a physical standpoint, the behavior of the gCg band reflects the underlying interpolation between a matter-dominated regime at high redshift and a DE-like phase at late times. At $z\sim2$, all curves converge to positive values of $q(z)$, consistent with decelerated expansion dominated by effective matter. Toward lower redshifts, the gCg band transitions smoothly to negative $q(z)$, reproducing the accelerated expansion regime. The fiducial gCg model deviates slightly in the transition region, yielding an earlier or less accurate crossing of $q=0$, which highlights the sensitivity of the model to the parameters $(\tilde{A},\alpha)$. On the other hand, the $z_t$-selected ensemble (2159 accepted models) tightly clusters around the $\Lambda$CDM prediction, showing that the apparent discrepancies often attributed to the gCg model are not intrinsic, but rather arise from unconstrained parameter choices. This result emphasizes that, when anchored to observationally motivated transition-redshift constraints derived from $H(z)$ data (cosmic chronometers, BAO, and DESI), the generalized Chaplygin gas remains a viable and competitive model at the background level.


\begin{figure}[h]
    \centering
    \includegraphics[width=0.50\textwidth]{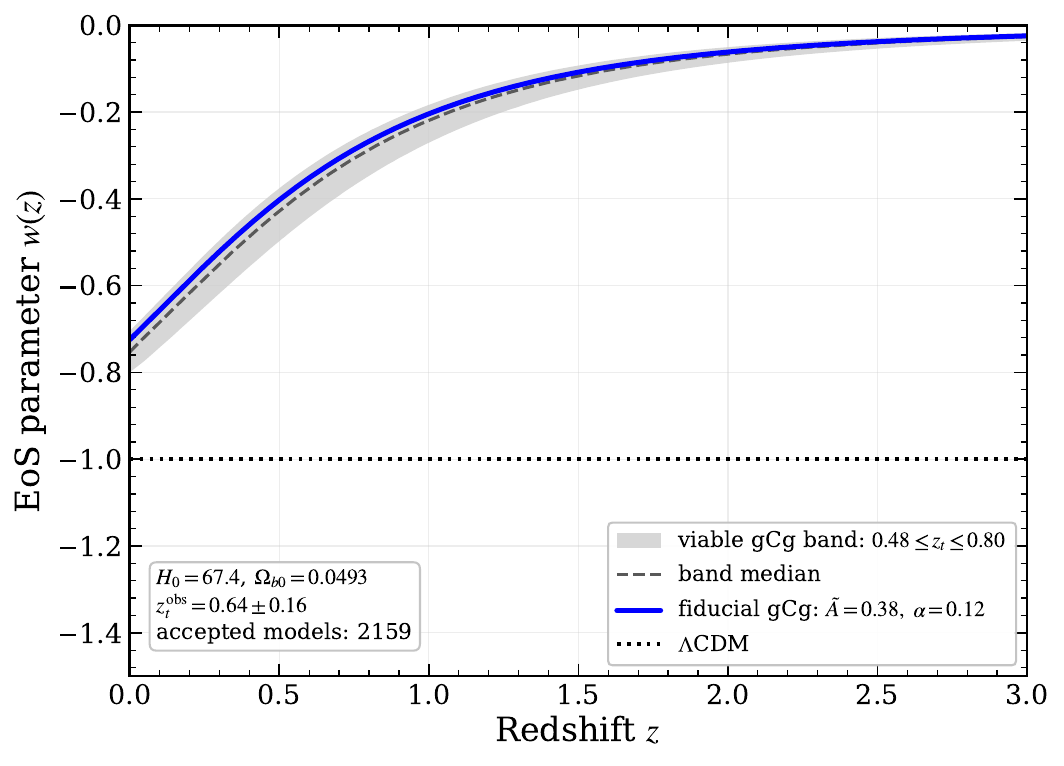}
    \caption{Evolution of the gCg EoS parameter $w_{\rm gCg}(z)$ in the redshift interval $0\leq z\leq3$. The gray band shows the viable gCg parameter region selected from the transition-redshift condition $0.48\leq z_t\leq0.80$, the gray dashed curve gives the median of this band, and the blue solid curve denotes the fiducial model $\tilde{A}=0.38$, $\alpha=0.12$. The black dotted line corresponds to the $\Lambda$CDM limit, $w=-1$. The gCg fluid evolves from a negative-pressure regime at low redshift toward a matter-like behavior, $w_{\rm gCg}\rightarrow0$, at higher redshift.}
    \label{w.UDM} 
    \end{figure}\par
\begin{figure}[h]
    \centering
    \includegraphics[width=0.50\textwidth]{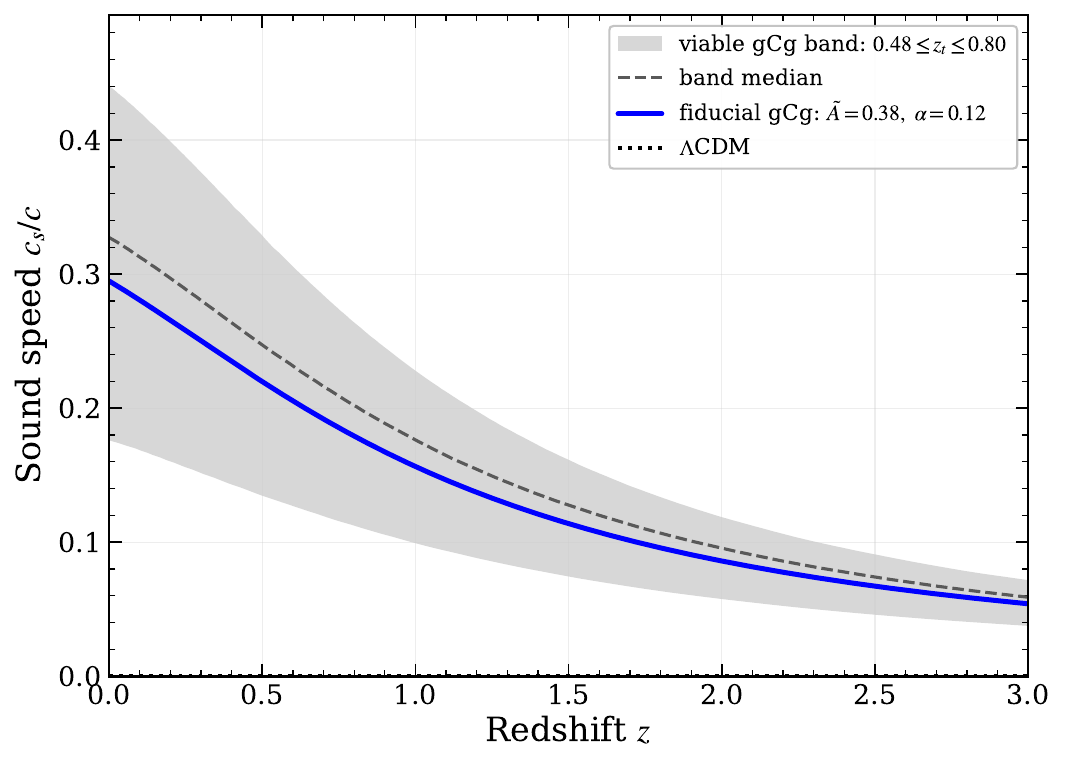}
    \caption{ Evolution of the adiabatic sound speed $c_s/c$ for the gCg model. The gray band represents the same transition-redshift-selected parameter region used in Fig.~\ref{w.UDM}, the gray dashed curve is the band median, and the blue solid curve corresponds to the fiducial model $\widetilde A=0.38$ and $\alpha=0.12$. The black dotted line, labeled
$\Lambda$CDM in the plot, denotes the pressureless cold-dark-matter reference limit, $c_{s,\mathrm{CDM}}/c=0$, rather than a sound speed related to the full $\Lambda$CDM dark sector. The cosmological constant does not describe a clustering fluid with propagating density perturbations.}
    \label{fig:sound.UDM} 
\end{figure}

Figure~\ref{w.UDM} shows that the transition-redshift-selected gCg band has the expected unified-fluid behavior. At low redshift the pressure is negative and drives accelerated expansion, whereas at increasing redshift the EoS approaches $w_{\rm gCg}\simeq0$, recovering an effectively matter-like phase. The fiducial curve $\tilde{A}=0.38$, $\alpha=0.12$ lies inside the viable band and therefore provides a representative background realization compatible with the transition-redshift constraint used in Fig.~\ref{fig:deceleration}. The comparison with the dotted $\Lambda$CDM reference emphasizes that the gCg does not behave as a pure cosmological constant; instead, it dynamically interpolates between DE-like and matter-like regimes.

One of the important questions to be answered in the Chaplygin models is related to the behavior of sound speed. In Fig.~\ref{fig:sound.UDM} , it shows the corresponding adiabatic sound speed, 
\begin{equation}\label{eq:sound_speed}
c_s^2=\frac{\partial p_{\rm gCg}}{\partial \rho_{\rm gCg}}
     =-\alpha w_{\rm gCg}\;.
\end{equation}
Since $w_{\rm gCg}<0$ at late times and $\alpha>0$, the adiabatic sound speed is positive and reaches its largest values at low redshift. Although these values are compatible with the background evolution and remain subluminal, they can generate pressure-gradient effects in the perturbation equations. In particular, a non-negligible adiabatic sound speed may suppress the growth of density perturbations or generate acoustic oscillations in the unified dark fluid. This sound-speed problem provides the motivation for the perturbative treatment developed in the Section \ref{sec:perturbations}. The background gCg model can be viable at the level of $H(z)$, $q(z)$, and $w(z)$, but a successful unified dark-sector model must also cluster consistently with large-scale-structure data. Therefore, before interpreting the gCg as a fully viable DM--DE unification, it is necessary to distinguish the background adiabatic sound speed from the effective rest-frame sound speed that controls pressure perturbations. The following section, we first review the adiabatic scalar perturbation equations and then we introduce the entropic/non-adiabatic prescription that allows us to get the phenomenologically important limit $c_{\rm eff}^2\simeq0$.

\section{Entropic perturbative regime}\label{sec:perturbations}

\subsection{Perturbed Einstein equations}

The perturbation equations in this section are presented in a way to motivate the distinction between the adiabatic sound speed $c_s^2$ and the effective rest-frame sound speed $c_{\rm eff}^2$. we point out that they are not the equations integrated in the numerical likelihood, which instead uses the reduced effective-growth prescription in Sec.~\ref{subsec:rsd_growth_twocol}. Actually, a central difficulty of unified Chaplygin-gas models arises at the
perturbative level. Although the background evolution naturally interpolates between a matter-like regime and a dark-energy-like phase, the adiabatic sound speed of \eqref{eq:sound_speed} becomes non-negligible at late times. The corresponding pressure-gradient
term is proportional to $c_s^2 k^2\delta/a^2$ and can generate acoustic oscillations, suppress the growth of density perturbations, and distort the matter power spectrum.

At small scales, as well known, CMB exhibits inhomogeneities in the distribution of matter, from which it is hypothesized that the structures present in the Universe come from the growth of initial small perturbations \cite{brandenberger}. The growth of structures is a consequence of gravitational instabilities, i.e., if there is a small fluctuation in the density of matter, this will lead to a point in space having a force of gravitational attraction slightly greater than at other points. If this were to occur in a scenario in which the universe is static, the result would be an instability that would project exponentially. However, in an expanding universe scenario, the gravitational instability is partially counterbalanced by the expansion itself \cite{mukhanov}. And this, in the long term of the universe evolution, would generate a linear regime of the perturbations. If we analyze the fluctuations in the CMB, we can see that they are on the order of $10^{-3}$ and this supports the idea of linear perturbations.

As the perturbations are related to fluctuations in the density of matter, then a perturbation in the stress tensor would generate perturbations in the Einstein tensor and, therefore, in the spacetime metric itself. The disturbance in the stress tensor is very small, so $|\delta T_{\mu\nu}|\ll |T_{\mu\nu}|$ and, consequently, $|\delta G_{\mu\nu}|\ll|G_{\mu\nu}|$ and $|\delta g_{\mu\nu}|\ll |g_{\mu\nu}|$.

According to \cite{mukhanov1} to describe cosmological perturbations for a dynamical expanding universe, we have to relate the background spacetime $\bar{g}_{\mu\nu}$ based on the flat FLRW metric with a perturbed spacetime $\delta g_{\mu\nu}$, in a way that the full perturbations evolves as $g_{\mu\nu}=\bar{g}_{\mu\nu}+\delta{g}_{\mu\nu}$. In the Newtonian gauge, the Bardeen potentials are characterized by the functions $ \Phi$ and $ \Psi$ and therefore the scalar-mode line element is given as follows
\begin{equation}
    ds^2=-a^2[(1+2\Phi)d\eta^2-(1-2\Psi)\delta_{ij}dx^idx^j]\;.
\end{equation}
The Einstein equation components in Fourier $k$-space are given by 
\begin{equation}
\left(\frac{k}{\mathcal{H}}\right)^2\Psi
=-\frac{3}{2}
\left[
\delta+3(1+w)\frac{\mathcal{H}}{k}v
\right].
\label{eq:einstein_constraint_corrected}
\end{equation}

\begin{equation}
 \label{poisson}
     k^2\Psi=-4\pi Ga^2\bar{\rho}_i\delta_i,
 \end{equation}
 
\begin{align}
\left(\frac{k}{\mathcal{H}}\right)^2(\Psi-\Phi)&=3w\Pi,&\\
\frac{\Psi'}{\mathcal{H}}+\Phi & =\frac{3}{2}(1+w)\frac{\mathcal{H}}{k}v,&\\
\frac{\Psi''}{\mathcal{H}^2}
&-\frac{1}{3}\left(\frac{k}{\mathcal{H}}\right)^2(\Phi-\Psi)
+\frac{1}{\mathcal{H}}(\Phi'+2\Psi')\nonumber\\
&+\left(1+\frac{2\mathcal{H}'}{\mathcal{H}^2}\right)\Phi
=\frac{3}{2}\frac{\delta p}{\bar\rho} .
\end{align}
where $\Pi$ refers to the anisotropic pressure along the fluid, $v$ is the fluid velocity, the dark fluid parameter denoted by $w=\frac{\bar{p}}{\bar{\rho}}$ and the fractional overdensity as $\delta=\frac{\delta \rho}{\bar{\rho}}$.

\subsection{Null entropy}
\label{subsec:null_entropy}

For completeness purposes, we first consider a perfect and adiabatic fluid (null entropy). In addition, we set $\Pi=0$ and for a perfect fluid $\Psi=\Phi$. Thus, the new set of equations is given by

\begin{equation}
\label{equação1}
    k^2\Phi=4\pi Ga^2 \bar{\rho}\left[\delta+3\mathcal{H}(1+w)\frac{\theta}{k^2}\right],
\end{equation}

\begin{equation}
\label{equação2}
    k^2\Phi-3\mathcal{H}(\Phi'+\mathcal{H}\Phi)=4\pi Ga^2\delta\rho,
\end{equation}

\begin{equation}
\label{equação3}
\Phi''+3\mathcal{H}\Phi'
+(\mathcal{H}^2+2\mathcal{H}')\Phi
=4\pi Ga^2\delta p
=4\pi Ga^2 c_s^2\delta\rho.
\end{equation}

\noindent where $\theta=\nabla_i v^i$ is the velocity divergence. For a single effective fluid, the conformal-Hubble relation is
\begin{equation}
\mathcal{H}'=-\frac{1}{2}\left(1+3w_{\rm eff}\right)\mathcal{H}^2,
\label{eq:Hconformal_corrected}
\end{equation}
and the comoving density perturbation is $\Delta=\delta+3\mathcal{H}(1+w)\theta/k^2$. The equations in the following are included only as theoretical motivation; the numerical likelihood does not integrate this full perturbation system. Combining the previous equations, and performing a transformation of variables $\eta \rightarrow t/a$, we find the following master equation for the evolution of matter overdensity as

\begin{equation}
\label{mestragCg}
\begin{aligned}
\ddot{\delta}
&+[1-3(2w_{gCg}-c_s^2)]\mathcal{H}\dot{\delta} \\
&-\frac{3}{2}(1-6c_s^2+8w_{gCg}-3w_{gCg}^2)
\mathcal{H}^2\delta
=-\left(\frac{k c_s}{a}\right)^2\delta .
\end{aligned}
\end{equation}

\noindent where $\dot{\delta}=d\delta/dt$, $w_{gCg}$ is the EoS given by (\ref{Eos gCg}) and $c_s$ is the speed of sound in the UDM model. Using $\xi\equiv\frac{\overset{\circ}{(H^2)}}{2H^2}=-{3}/{2}[1+(\frac{A+B}{B}-1)a^{3(1+\alpha)}]^{-1}$, we can rewrite equation (\ref{mestragCg}) conveniently as

\begin{equation}
\begin{aligned}
\overset{\circ\circ}{\delta}
&+[1+\xi-3(2w_{gCg}-c_s^2)]\overset{\circ}{\delta} \\
&-\frac{3}{2}(1-6c_s^2+8w_{gCg}-3w_{gCg}^2)\delta
=-\left(\frac{k c_s}{a}\right)^2\delta .
\end{aligned}
\end{equation} 
This differential equation has two solutions, a decreasing $D_-$ and an increasing $D_+$. We are interested in the $D_+$ growth mode (since the other loses importance over time), so we can relate this mode to the density contrast such that 

\begin{equation}
\delta(t)=\frac{D_+(t)}{D_+(t_i)}\delta(t_i), 
\end{equation}

\noindent where $D_+(t_i)$ and $\delta(t_i)$ are constants given by the initial conditions. Unfortunately, the adiabatic perturbative solutions are not shown as final viability results,
since the large adiabatic sound speed produces scale-dependent oscillations
and suppresses structure growth as shown in Fig.\eqref{fig:sound.UDM}. Instead, they should be interpreted only as
a diagnostic reference case motivating in the following the non-adiabatic prescription
$c_{\rm eff}^2\simeq0$.

\subsection{Entropic Chaplygin gas}
\label{subsec:entropic_cg}

The total entropy perturbation is defined as

\begin{equation}
\mathcal{S} = \mathcal{H}\left( \frac{\delta p}{\bar{p}'}- \frac{\delta \rho}{\bar{\rho}'}\right),
\end{equation}

\noindent which is gauge invariant. From background relations $\bar{\rho}' = -3\mathcal{H}(1+w_{gCg})\bar{\rho}$ and $\bar{p}'=c_s^2 \bar{\rho}'$, the last equation becomes

\begin{equation}
\mathcal{S} = \frac{1}{3(1+w_{gCg})}\left( \delta- \frac{\delta p}{c_s^2\bar{\rho}}\right),
\end{equation}

\noindent which yields 

\begin{equation}
    \delta p = c_s^2\bar{\rho} [\delta - 3(1+w_{gCg})\mathcal{S}].
\end{equation}

The pressure perturbation above shows explicitly how the entropy mode can change the clustering properties of the gCg fluid without changing its background EoS. Therefore, we define an effective rest-frame sound speed by $c_{\rm eff}^2\equiv \delta p_{\rm eff}/\delta\rho$. The phenomenologically viable cold-clustering limit is obtained by imposing $c_{\rm eff}^2\simeq0$, or equivalently $\delta p_{\rm eff}\simeq0$. In the present notation this corresponds to
\begin{equation}
\mathcal{S}=\frac{\delta}{3(1+w_{gCg})},
\end{equation}
which cancels the adiabatic pressure perturbation and removes the pressure-gradient contribution responsible for acoustic oscillations in the matter overdensity. Thus $c_s^2$ remains the background adiabatic sound speed, while $c_{\rm eff}^2\simeq0$ is the perturbative prescription required for CDM-like growth of structures. And, substituting the pressure perturbation relation in Eq.\ref{equação3}, we get

\begin{equation}
\begin{aligned}
\Phi''+3\mathcal{H}\Phi'
+(\mathcal{H}^2+2\mathcal{H}')\Phi
&=4\pi Ga^2 c_s^2\delta\rho \\
&-12\pi Ga^2 c_s^2\bar\rho(1+w_{gCg})\mathcal{S} .
\end{aligned}
\end{equation}
Hence the adiabatic pressure-gradient contribution is canceled by the non-adiabatic entropy contribution rather than added to it. It is worth noting that we do not retain a separate expanded ``master equation'' for the entropy term, whose sign and normalization are convention dependent. In the following likelihood analysis, we do not integrate the complete scale-dependent unified-fluid perturbation system or the full Einstein--Boltzmann hierarchy. Instead, for now, we adopt the reduced sub-horizon effective-growth prescription specified in Eqs.~\eqref{eq:growth_twocol}--\eqref{eq:Omega_gcg_m_twocol}.

\section{Datasets, likelihoods, and numerical implementation}
\label{sec:data_likelihoods}

The constraints reported in this work are obtained with the likelihood structure implemented in the accompanying analysis code. We consider Planck+DESI+RSD+PPS and, replacing PPS by the DES--Dovekie relative Hubble diagram, Planck+DESI+RSD+DES--Dovekie. Moreover, we use the Planck compressed quantities that follow the 2018 Planck cosmological-parameter analysis and the associated distance-prior construction \cite{ChenHuangWangPlanckDistancePriors,Planck2020Parameters}. The BAO likelihood uses the DESI DR2 Gaussian BAO vector and covariance \cite{DESIDR2BAO}, and for the supernova data,  our implementation uses PantheonPlus+SH0ES \cite{Scolnic2022PantheonPlusData,Brout2022PantheonPlus} and DES--Dovekie \cite{Popovic2025Dovekie} public data. The RSD term follows the standard use of $f\sigma_8$ growth-rate measurements \cite{Beutler2012RSD,Howlett2015RSD,eBOSS2021}.

The joint likelihood is written as a product of independent Gaussian terms.  Equivalently,
\begin{equation}
\begin{split}
 \chi^2_{\rm tot} ={}& \chi^2_{\rm SN}
 + \chi^2_{\rm DESI}
 + \chi^2_{\rm RSD}
 + \chi^2_{\rm Pl,dist} + \chi^2_{\rm Pl,lens}
\end{split}
\label{eq:total_chi2_twocol}
\end{equation}
with $\ln {\cal L}_{\rm tot}=-\chi^2_{\rm tot}/2$.  In our implementation, the BBN and external $S_8$ Gaussian priors are not used.  Thus, no artificial Gaussian prior is imposed on $S_8$ and
\begin{equation}
 S_8=\sigma_{8,0}\left(\frac{\Omega_m}{0.3}\right)^{1/2}
\label{eq:S8_twocol}
\end{equation}
is computed as a derived quantity.  The numerical constants used in the likelihood evaluation are $T_{\rm CMB}=2.7255\,{\rm K}$, $N_{\rm eff}=3.046$, and $c=299792.458\,{\rm km.s^{-1}}$.

\subsection{Models and sampled parameters}
\label{subsec:model_parameters_twocol}
Once the entropic UDM parameter space can be non-Gaussian and strongly degenerate,  we use \texttt{MultiNest}~\cite{Feroz2009MultiNest,Feroz2019,Buchner2014PyMultiNest} to explore better the posterior and compute the Bayesian evidence. This sampling choice is logically distinct from an Einstein--Boltzmann solver such as\texttt{CLASS}~\cite{Blas:2011rf} or \texttt{CAMB}~\cite{Lewis:1999bs} that would be required to predict full CMB spectra after the entropic perturbation sector has been consistently implemented, whereas \texttt{MultiNest} performs inference using the observables supplied by the present background and effective-growth calculation.

In the following, we present the necessary change of definition of some terms and parameters that are conveniently written in the code. We start with the entropic UDM Chaplygin-gas model which is sampled with
\begin{equation}\label{eq:priors}
 \boldsymbol{\theta}_{\rm gCg}=
 \left(H_0,\Omega_b,\widetilde A,\alpha,\sigma_{8,0}\right),
\end{equation}
where the flat priors implemented in \texttt{MultiNest} code. We use prior transform are $50<H_0<90$, $0.03<\Omega_b<0.08$, $0.05<\widetilde A<5.0$,$0\leq\alpha\leq1$,$
0.45<\sigma_{8,0}<1.20$. It is worth noting that the code uses $\widetilde A=B/A$, so the usual normalized Chaplygin-gas parameter is
\begin{equation}
 B_s=\frac{1}{1+\widetilde A}.
\end{equation}
The normalized gCg density evolution is
\begin{equation}
 F_{\rm gcg}(a)=
 \left[
 \frac{1+\widetilde A a^{-3(1+\alpha)}}{1+\widetilde A}
 \right]^{1/(1+\alpha)},
 \label{eq:Fgcg_twocol}
\end{equation}
while its background EoS and adiabatic sound speed are now given by
\begin{equation}
 w_{\rm gcg}(a)=
 -\frac{1}{1+\widetilde A a^{-3(1+\alpha)}},
 \quad
 c_{s,{\rm gcg}}^2(a)=-\alpha w_{\rm gcg}(a).
 \label{eq:wcs_gcg_twocol}
\end{equation}
Spatial flatness fixes the cosmological parameter related to Chapligyn gas as $\Omega_{g,0}=1-\Omega_b-\Omega_r$.  We also define the effective present matter density used for comparison with $\Lambda$CDM as
\begin{equation}
 \Omega_m=\Omega_b+
 \Omega_{g,0}
 \left(\frac{\widetilde A}{1+\widetilde A}\right)^{1/(1+\alpha)}.
 \label{eq:Omega_m_gcg_twocol}
\end{equation}
Then, the gCg background expansion is simply given by
\begin{equation}
 E^2(a)=\Omega_b a^{-3}+\Omega_r a^{-4}
 +\Omega_{g,0}F_{\rm gcg}(a).
 \label{eq:E2_gcg_twocol}
\end{equation}

The flat $\Lambda$CDM baseline is sampled with
\begin{equation}
 \boldsymbol{\theta}_{\Lambda{\rm CDM}}=
 \left(H_0,\Omega_b,\Omega_{\rm cdm},\sigma_{8,0}\right),
\end{equation}
using the same prior ranges for $H_0$, $\Omega_b$, and $\sigma_{8,0}$, with $\Omega_{\rm cdm}$ replacing the gCg sector parameters.  In this case we have the standard expressions $\Omega_m=\Omega_b+\Omega_{\rm cdm}$, $\Omega_\Lambda=1-\Omega_m-\Omega_r$, and the $\Lambda$CDM background expansion as 
\begin{equation}
 E^2(a)=\Omega_m a^{-3}+\Omega_r a^{-4}+\Omega_\Lambda .
 \label{eq:E2_lcdm_twocol}
\end{equation}

\subsection{Distances and sound horizon}
\label{subsec:distances_twocol}

All background observables are computed from the same expansion function.  Assuming zero spatial curvature, the transverse comoving distance is
\begin{equation}
 D_M(z)=\frac{c}{H_0}\int_0^z\frac{dz'}{E(z')},
 \label{eq:DM_twocol}
\end{equation}
and the luminosity distance and supernova distance modulus are
\begin{equation}
 D_L(z)=(1+z)D_M(z),
 \qquad
 \mu_{\rm th}=5\log_{10}\!\left(\frac{D_L}{\rm Mpc}\right)+25 .
 \label{eq:mu_twocol}
\end{equation}
The baryon drag redshift entering the DESI BAO prediction is computed using the Eisenstein--Hu fitting relation \cite{EisensteinHu1998},
\begin{equation}
 z_d =
 \frac{1291(\Omega_m h^2)^{0.251}}
 {1+0.659(\Omega_m h^2)^{0.828}}
 \left[1+b_1(\Omega_b h^2)^{b_2}\right],
 \label{eq:zdrag_twocol}
\end{equation}
where
\begin{align}
b_1 &= 0.313\left(\Omega_m h^2\right)^{-0.419}
\left[1+0.607\left(\Omega_m h^2\right)^{0.674}\right],\nonumber
\\
b_2 &= 0.238\left(\Omega_m h^2\right)^{0.223}.
\end{align}
The sound horizon at drag is evaluated internally as
\begin{equation}
 r_d=\int_0^{a_d}\frac{c_s(a)}{a^2H(a)}\,da,
 \qquad a_d=(1+z_d)^{-1},
 \label{eq:rd_twocol}
\end{equation}
with
\begin{equation}
 c_s(a)=\frac{c}{\sqrt{3[1+R(a)]}},
 \qquad
 R(a)=\frac{3\Omega_b}{4\Omega_\gamma}a.
\end{equation}
The photon-decoupling redshift used in the Planck distance-prior sector follows the Hu--Sugiyama fitting form \cite{HuSugiyama1996}.

\subsection{Supernova likelihoods}
\label{subsec:sn_likelihoods_twocol}

For Planck+DESI+RSD+PPS, we use the official 1701-object PantheonPlus+SH0ES data vector and its full statistical-plus-systematic covariance matrix. The sample contains 77 Cepheid-host calibrator entries. 
Let $\mathcal{C}$ and $\mathcal{H}$ denote, respectively, the
Cepheid-calibrator and Hubble-flow subsets, for each supernova we define
\begin{align}
m^{\rm mod}_{B,i}(M_B)
&= M_B+\mu_i^{\rm SN},
\label{eq:pps_model_twocol}
\\
\mu_i^{\rm SN}
&=
\begin{cases}
\mu_{{\rm Ceph},i},
    & i\in\mathcal{C},\\[2pt]
\mu_{\rm th}(z_{{\rm HD},i},z_{{\rm hel},i}),
    & i\in\mathcal{H}.
\end{cases}
\label{eq:pps_mu_twocol}
\end{align}
where $\mu_{{\rm Ceph},i}$ is the supplied \texttt{CEPH\_DIST} anchor and the non-calibrator prediction uses the supplied Hubble-diagram and heliocentric redshifts. With $\Delta{\bf m}={\bf m}_{B,\rm corr}-{\bf m}^{\rm mod}_{B}$, the likelihood is
\begin{equation}
 \chi^2_{\rm PPS}(M_B)=
 \Delta{\bf m}^{T}{\bf C}_{\rm PPS}^{-1}\Delta{\bf m}.
\label{eq:chi2_pps_twocol}
\end{equation}
The common absolute magnitude $M_B$ is profiled analytically at every cosmological point and is counted as one fitted nuisance parameter in AIC, BIC, and the number of degrees of freedom. We point out that the same vector, covariance, calibrator mask, redshift convention, and nuisance treatment are used for gCg and $\Lambda$CDM. Including the Cepheid-host branch is essential because it retains the absolute calibration information carried by PPS, whereas the non-calibrator branch constrains the shape of the Hubble diagram.

On the other hand, for DES-Dovekie data, the full inverse covariance is reconstructed, mirrored across the diagonal, and symmetrized as
\begin{equation}
 {\bf W}_{\rm DESD}=\frac{1}{2}
 \left({\bf C}_{\rm DESD}^{-1}+{\bf C}_{\rm DESD}^{-T}\right).
\end{equation}
In the loaded run the matrix has dimension $1820\times1820$ and diagonal range
\begin{equation}
 4.565499\times10^{-6}
 \leq {\rm diag}({\bf W}_{\rm DESD})
 \leq 2.536374\times10^{2}.
\end{equation}
We point out that DES-Dovekie is treated as a relative Hubble diagram, then the additive distance-modulus offset is analytically marginalized.  Therefore, defining $\Delta\boldsymbol{\mu}=\boldsymbol{\mu}_{\rm obs}-\boldsymbol{\mu}_{\rm th}$ and ${\bf 1}=(1,\ldots,1)^T$, the implemented quantities are
\begin{align}
 C &= \Delta\boldsymbol{\mu}^{T}
      {\bf W}_{\rm DESD}
      \Delta\boldsymbol{\mu},
 \nonumber\\
 B &= {\bf 1}^{T}
      {\bf W}_{\rm DESD}
      \Delta\boldsymbol{\mu},
 \nonumber\\
 A &= {\bf 1}^{T}
      {\bf W}_{\rm DESD}
      {\bf 1},
 \label{eq:abc_desd}
\end{align}
and the marginalized chi-square is
\begin{equation}
 \chi^2_{\rm DESD}
 = C-\frac{B^2}{A}.
 \label{eq:chi2_desd_twocol}
\end{equation}
This expression is the Gaussian result for profiling or marginalizing over a constant nuisance offset with a flat prior. We count that offset as one fitted nuisance parameter, consistently with the PPS treatment.

\subsection{DESI DR2 BAO likelihood}
\label{subsec:desi_bao_twocol}

The BAO likelihood uses the DESI DR2 Gaussian BAO mean vector and covariance.  The code reads the redshift, observed value, and observable label corresponding to
$\frac{D_M(z)}{r_d}$, $\frac{D_H(z)}{r_d}$, $\frac{D_V(z)}{r_d}$. Here $D_H(z)=c/H(z)$ and
\begin{equation}
 D_V(z)=\left[D_M^2(z)\frac{cz}{H(z)}\right]^{1/3}.
 \label{eq:DV_twocol}
\end{equation}
For a theoretical vector ${\bf y}_{\rm th}$ and observed DESI vector ${\bf y}_{\rm DESI}$, the Gaussian term is
\begin{equation}
 \chi^2_{\rm DESI}=({\bf y}_{\rm DESI}-{\bf y}_{\rm th})^T
 {\bf C}_{\rm DESI}^{-1}({\bf y}_{\rm DESI}-{\bf y}_{\rm th}),
 \label{eq:chi2_desi_twocol}
\end{equation}
with the covariance symmetrized and Cholesky-factorized before evaluation.

\subsection{Planck compressed CMB likelihoods}
\label{subsec:planck_twocol}

The CMB information is implemented through Planck 2018 compressed distance priors and an additional Planck lensing-amplitude prior.  The distance-prior vector is
\begin{equation}
 {\bf x}_{\rm Pl}=\left(R,\ell_A,\Omega_bh^2\right),
 \label{eq:planck_vector_twocol}
\end{equation}
where
\begin{equation}
 R=\sqrt{\Omega_m}\frac{H_0D_M(z_\star)}{c},
 \qquad
 \ell_A=\pi\frac{D_M(z_\star)}{r_s(z_\star)}.
\end{equation}
The numerical mean vector used in the code is
\begin{equation}
 \bar{\bf x}_{\rm Pl}=\left(1.750235,\,301.4707,\,0.02235976\right),
\end{equation}
and the inverse covariance is
\begin{equation}
\scriptsize
 {\bf C}_{\rm Pl}^{-1}=
 \begin{pmatrix}
 94392.3971 & -1360.4913 & 1664517.2916\\
 -1360.4913 & 161.4349 & 3671.6180\\
 1664517.2916 & 3671.6180 & 79719182.5162
 \end{pmatrix}.
 \label{eq:planck_invcov_twocol}
\end{equation}
\normalsize
Thus,
\begin{equation}
 \chi^2_{\rm Pl,dist}=({\bf x}_{\rm Pl}-\bar{\bf x}_{\rm Pl})^T
 {\bf C}_{\rm Pl}^{-1}({\bf x}_{\rm Pl}-\bar{\bf x}_{\rm Pl}).
 \label{eq:chi2_planck_dist_twocol}
\end{equation}
The Planck lensing contribution is imposed as a Gaussian constraint on $\sigma_8\Omega_m^{1/4}$,
\begin{equation}
 \chi^2_{\rm Pl,lens}=\left[
 \frac{\sigma_{8,0}\Omega_m^{1/4}-0.589}{0.020}
 \right]^2 .
 \label{eq:chi2_planck_lens_twocol}
\end{equation}

An important note deserves attention. The distance-prior likelihood retains the dominant geometrical information contained in the acoustic scale, shift parameter, and physical baryon density, but it is not equivalent to the full Planck likelihood. A full TT, TE, EE, and lensing analysis would additionally constrain the relative acoustic-peak heights, detailed spectral shapes, early and late integrated Sachs--Wolfe contributions, evolution of the gravitational potentials, and lensing smoothing. These observables depend on the perturbation dynamics of the unified fluid and are not completely specified by its background EoS. Moreover, full CMB information would be expected to reduce the allowed parameter volume, particularly along the degeneracies involving $\widetilde A$, $\alpha$, $H_0$, and the clustering amplitude, and could shift the inferred values of $H_0$, $\sigma_8$, and $S_8$. It could also change the Bayesian evidence substantially once if the entropic-gCg spectra remain close to those of $\Lambda$CDM, the main effect would likely be tighter constraints, whereas modifications of the acoustic peaks, gravitational potentials, or lensing spectrum could shift the posterior or exclude part of the parameter region allowed by the distance prior. The direction and magnitude of these changes cannot be determined reliably without calculating the complete CMB spectra.

Our analysis employs the gCg background expansion together with a phenomenological sub-horizon growth equation and the cold-clustering prescription $c_{\rm eff}^{2}\simeq0$. A consistent full-CMB calculation additionally requires evolution equations for the unified-fluid density and velocity perturbations, a covariant prescription for the non-adiabatic pressure perturbation and entropy mode, physically consistent initial conditions, and stable evolution toward the $w\rightarrow-1$ regime. Consequently, the present work is intentionally restricted to a background-plus-effective-growth test and does not constitute a full-CMB validation of the model that will be tackled in a future work.

\subsection{RSD likelihood and growth calculation}\label{subsec:rsd_growth_twocol}
\label{subsec:rsd_twocol}

The RSD module uses the observable $f\sigma_8(z)$, with
\begin{equation}
 f(z)=\frac{d\ln D}{d\ln a},\qquad
 \sigma_8(z)=\sigma_{8,0}D(z),\qquad
 f\sigma_8=f\sigma_8(z).
\end{equation}
The linear growth factor is obtained by integrating the reduced sub-horizon equation
\begin{equation}
 \frac{d^2D}{d\ln a^2}
 +\left[2+\frac{d\ln H}{d\ln a}\right]\frac{dD}{d\ln a}
 -\frac{3}{2}\Omega_{\rm cl}(a)D=0,
 \label{eq:growth_twocol}
\end{equation}
from $a_{\rm ini}=10^{-3}$ to $a=1$, with initial conditions $D(a_{\rm ini})=a_{\rm ini}$ and $dD/d\ln a|_{a_{\rm ini}}=a_{\rm ini}$. The solution is normalized to $D(1)=1$. In the effective entropic-gCg prescription implemented in the numerical analysis, the clustering source is
\begin{equation}
 \Omega_{\rm cl}(a)=\Omega_b(a)+\Omega_{\rm gcg,m}(a),
 \label{eq:Omega_cl_twocol}
\end{equation}
where we denote
\begin{align}
 \qquad\Omega_{\rm gcg,m}(a)
 &=
 \frac{\Omega_{\rm gcg,m,0}\,a^{-3}}{E^2(a)},
 \nonumber\\
 \Omega_{\rm gcg,m,0}&=
 \Omega_{\rm gcg,0}
 \left(
 \frac{\widetilde A}{1+\widetilde A}
 \right)^{1/(1+\alpha)}.
 \label{eq:Omega_gcg_m_twocol}
\end{align}
Thus, only the effective matter-like contribution of the unified gCg sector enters the source term. This construction guarantees that the $\alpha=0$ limit reproduces the standard flat-$\Lambda$CDM clustering equation. For $\alpha\neq0$, Eq.~\eqref{eq:growth_twocol} is used as a phenomenological effective-growth approximation associated with the cold-clustering condition $c_{\rm eff}^2\simeq0$; it is not a full Einstein--Boltzmann evolution of the unified fluid.

\begin{table}[t]
\centering
\caption{RSD growth-rate data used in our numerical analysis.}
\label{tab:rsd_main_twocol}
\begin{tabular}{ccc}
\toprule
$z$ & $f\sigma_8$ & $\sigma$\\
\midrule
0.067 & 0.423 & 0.055\\
0.150 & 0.490 & 0.145\\
0.380 & 0.497 & 0.045\\
0.510 & 0.458 & 0.038\\
0.610 & 0.436 & 0.034\\
0.700 & 0.437 & 0.034\\
\bottomrule
\end{tabular}
\end{table}
In Table \eqref{tab:rsd_main_twocol} we show the RSD data used for the runs. Since no off-diagonal RSD covariance is supplied in the selected run configuration, we have
\begin{equation}
 \chi^2_{\rm RSD}=\sum_i
 \left[\frac{(f\sigma_8)_{{\rm obs},i}-(f\sigma_8)_{{\rm th},i}}{\sigma_i}\right]^2 .
 \label{eq:chi2_rsd_twocol}
\end{equation}

\subsection{MultiNest sampling and posterior analysis}
\label{subsec:multinest_twocol}

The posterior exploration is performed with \texttt{MultiNest}, accessed from Python through \texttt{pymultinest} \cite{Skilling2006NestedSampling,Feroz2009MultiNest,Feroz2019,Buchner2014PyMultiNest}.  The sampler receives two functions from the code: a prior transform, which maps a unit hypercube vector to the physical parameter ranges in Eq.\eqref{eq:priors}, and a log-likelihood function,
\begin{equation}
 \ln {\cal L}(\boldsymbol{\theta})=-\frac{1}{2}\chi^2_{\rm tot}(\boldsymbol{\theta}).
\end{equation}
Parameter points producing invalid backgrounds, non-finite distances, negative densities, failed growth integrations, or non-finite likelihood contributions are rejected by returning an effectively vanishing likelihood.  Before a forced clean run, we defined a helper routine that removes existing files with the same basename, ensuring that the posterior sample and evidence estimate are not mixed with an earlier run.  

The \texttt{MultiNest} options used in the production runs are summarized in Table~\ref{tab:multinest_options_twocol}. All four model--dataset combinations use 800 live points and an evidence tolerance of $10^{-3}$. Multimodal mode and importance nested sampling are enabled, constant-efficiency mode is disabled, and the random seed is fixed to 123456. Posterior summaries are read with \texttt{pymultinest.Analyzer}, while GetDist is used to project the chains onto the common derived basis $\{H_0,\Omega_m,S_8\}$.

\begin{table}[t]
\centering
\caption{MultiNest settings used in all valid runs.}
\label{tab:multinest_options_twocol}
\begin{tabular}{lc}
\toprule
Option & Value\\
\midrule
\texttt{evidence\_tolerance} & 0.001\\
\texttt{sampling\_efficiency} & 0.35\\
\texttt{importance\_nested\_sampling} & True\\
\texttt{multimodal} & True\\
\texttt{const\_efficiency\_mode} & False\\
\texttt{n\_live\_points} & 800\\
\texttt{n\_iter\_before\_update} & 800\\
\texttt{seed} & 123456\\
\bottomrule
\end{tabular}
\end{table}

For model comparison, the generalized Chaplygin-gas samples are transformed by computing $\Omega_m$ from Eq.~\eqref{eq:Omega_m_gcg_twocol} and $S_8$ from Eq.~\eqref{eq:S8_twocol}.  The $\Lambda$CDM chains use $\Omega_m=\Omega_b+\Omega_{\rm cdm}$ and the same $S_8$ definition. It is worth noting that the information-criterion and tension-analysis cells additionally use the minimum chain value of $-2\ln {\cal L}$, the approximate number of data points, and the projected posterior covariance matrices.

\section{Numerical results and tension diagnostics}\label{sec:tension_diag}

\begin{figure}
    \centering
    \includegraphics[width=1\linewidth]{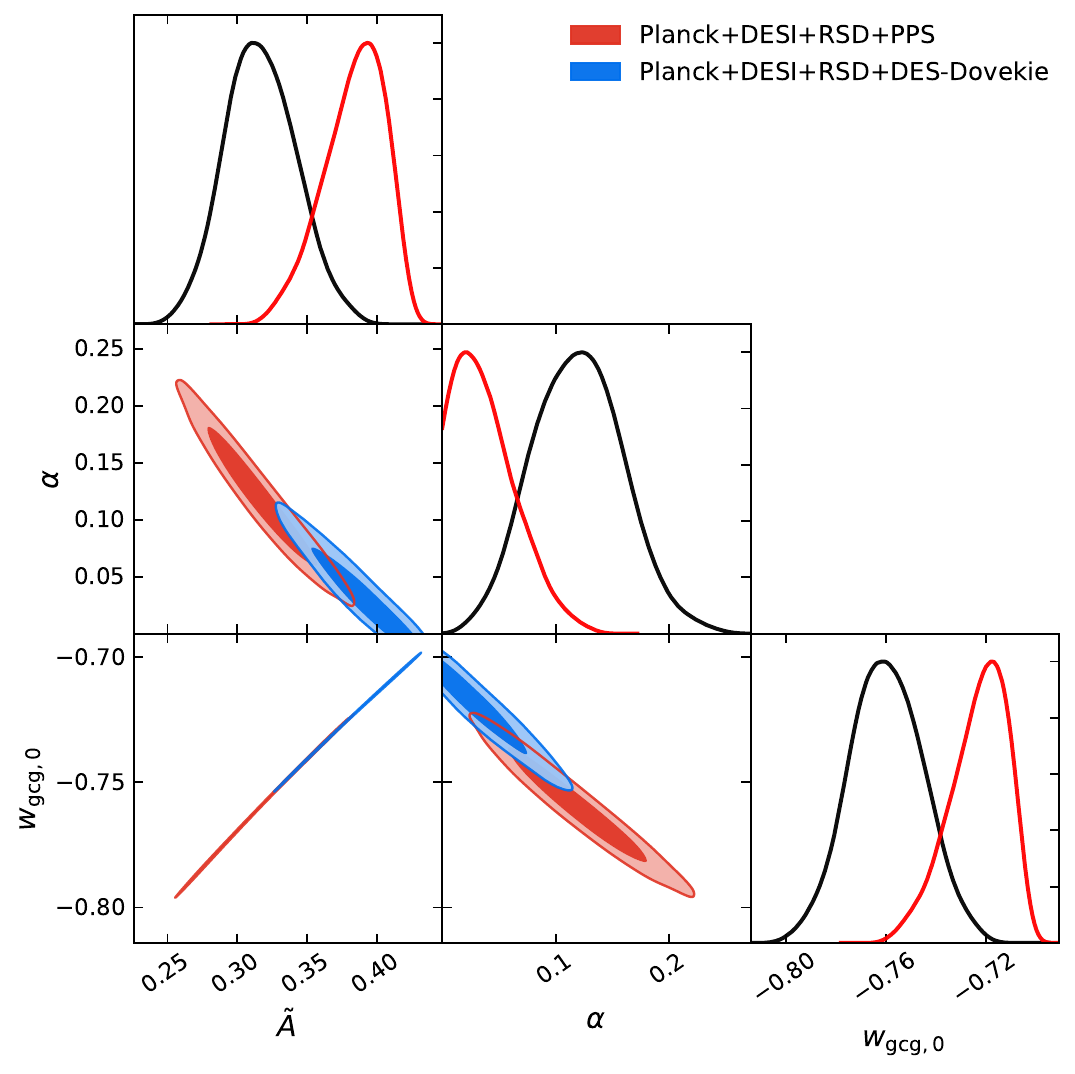}
    \caption{Posterior constraints on the entropic-gCg parameters for the two supernova combinations. Red contours show Planck+DESI+RSD+PPS and blue contours show Planck+DESI+RSD+DES--Dovekie. The parameters are the transition parameter $\widetilde A$, the Chaplygin exponent $\alpha$, and the derived present-day EoS $w_{{\rm gcg},0}$. PPS favours a larger $\alpha$ and, because $w_{{\rm gcg},0}=-1/(1+\widetilde A)$, a more negative present-day EoS. The DES--Dovekie posterior lies closer to the $\alpha=0$ background limit.}    \label{fig:gCg_parameters}
\end{figure}

In this section, we propose a numerical analysis to test the entropic generalized Chaplygin-gas model at two complementary levels. First, the posterior distributions determine whether the model reproduces the background expansion constrained by supernovae, DESI BAO, RSD, and compressed Planck information. Second, the chains are projected onto the common derived basis $\{H_0,\Omega_m,S_8\}$, permitting a direct comparison with $\Lambda$CDM and separating background-sensitive shifts from growth-sensitive behaviour.

\begin{figure*}
    \centering
    \includegraphics[width=1\linewidth]{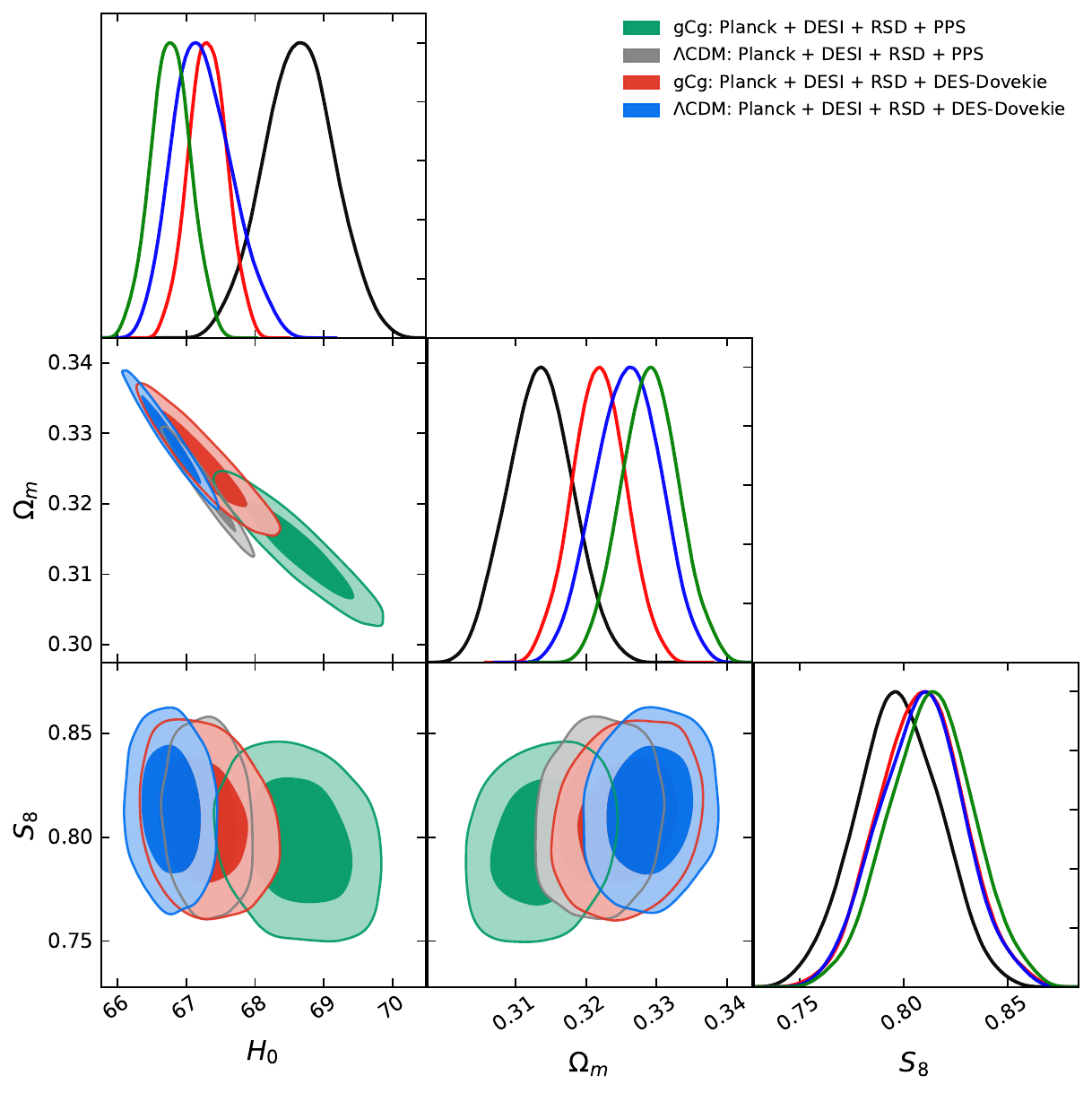}
    \caption{Comparison between entropic gCg and $\Lambda$CDM in the derived space $\{H_0,\Omega_m,S_8\}$. Green and gray contours denote gCg and $\Lambda$CDM with Planck+DESI+RSD+PPS; red and blue denote the corresponding DES--Dovekie combinations. The green PPS+gCg contour is displaced toward simultaneously larger $H_0$ and smaller $S_8$ and is the most favourable of the four cases for a joint partial alleviation of the two tensions. Its marginalized constraints are $H_0=68.64\pm0.52$, $\Omega_m=0.3135\pm0.0047$, and $S_8=0.797\pm0.021$, compared with $H_0=67.31\pm0.28$, $\Omega_m=0.3218\pm0.0039$, and $S_8=0.809\pm0.021$ for PPS+$\Lambda$CDM. The shift remains modest relative to the local $H_0$ discrepancy, and the substantial $S_8$ overlap means that the figure supports partial alleviation. Quoted errors are marginalized 68\% credible intervals.}
    \label{fig:triang_plots}
\end{figure*}

\begin{figure*}[t]
    \centering
    \begin{subfigure}[t]{0.48\linewidth}
        \centering
        \includegraphics[width=\linewidth]{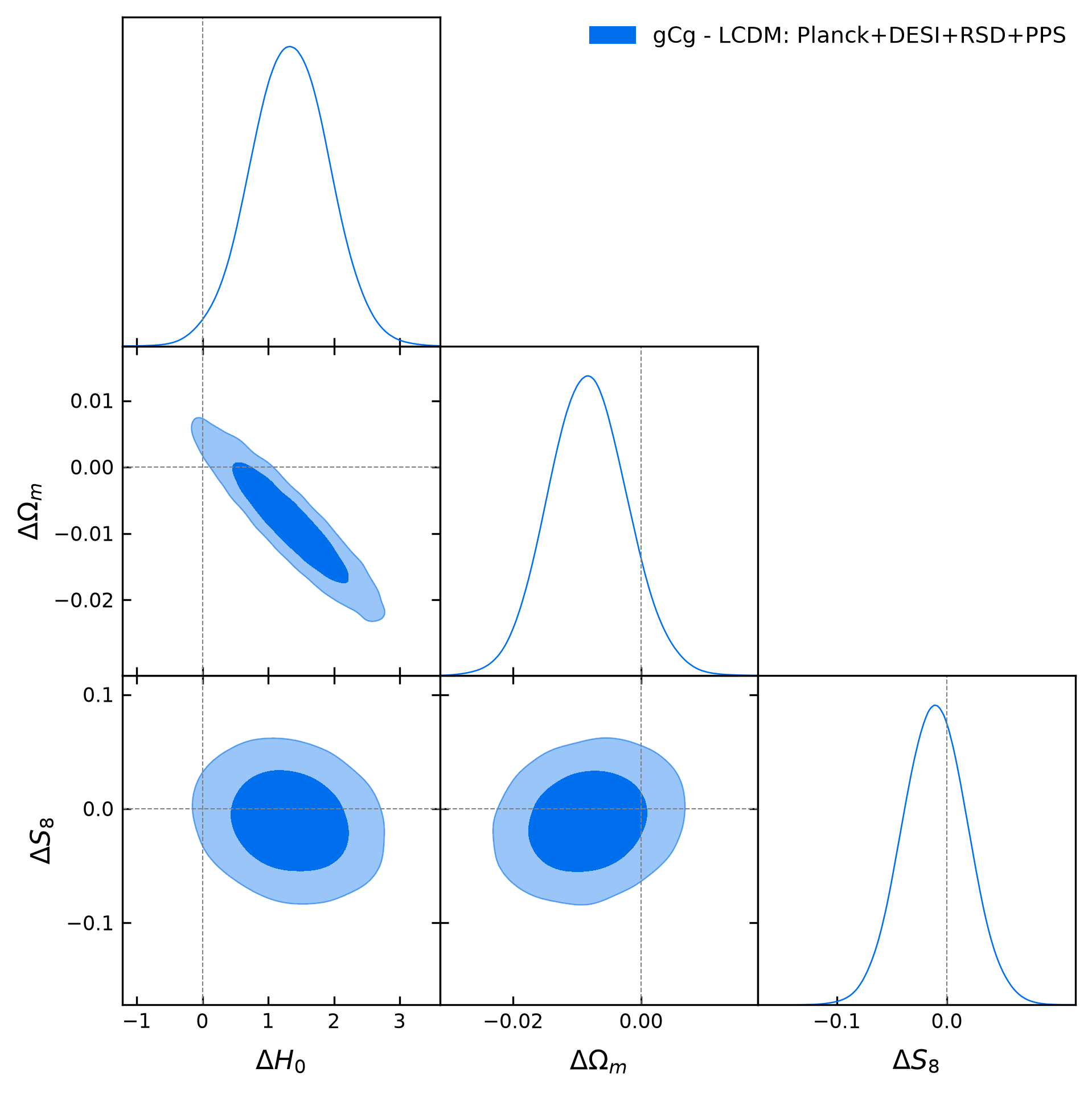}
        \caption{PPS configuration.}
        \label{fig:tens_pps}
    \end{subfigure}%
    \hfill
    \begin{subfigure}[t]{0.48\linewidth}
        \centering
        \includegraphics[width=\linewidth]{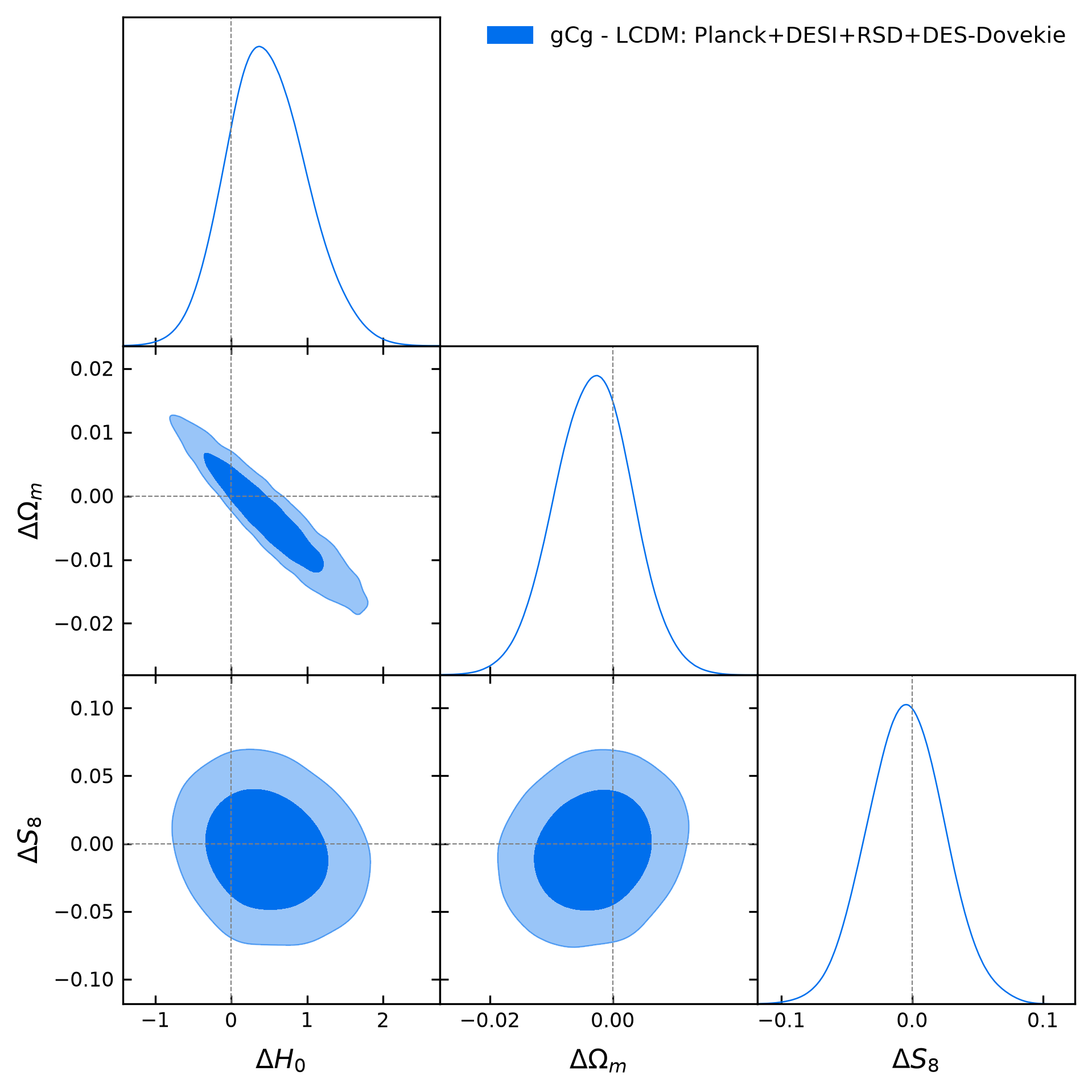}
        \caption{DES--Dovekie configuration.}
        \label{fig:tens_desd}
    \end{subfigure}
    \caption{Posterior-difference distributions, gCg minus $\Lambda$CDM, evaluated with \texttt{Tensiometer} in the common derived space $(H_0,\Omega_m,S_8)$ for PPS (left) and DES--Dovekie (right). The PPS distribution is centered near $(\Delta H_0,\Delta\Omega_m,\Delta S_8)\simeq(1.33,-0.0083,-0.011)$ and the DES--Dovekie distribution is centered near $(0.44,-0.0031,-0.0046)$. The signs of the PPS shifts are simultaneously favourable for reducing the $H_0$ and $S_8$ discrepancies. These are cross-model posterior displacements obtained from the same joint dataset, not direct early-versus-late tension significances.}
    \label{fig:tens2}
\end{figure*}

In Figure~\ref{fig:gCg_parameters}, we have a comparison of the gCg constraints with the two supernova combinations. PPS gives $\widetilde A=0.316^{+0.028}_{-0.025}$ and $\alpha=0.120^{+0.041}_{-0.042}$, whereas DES--Dovekie gives $\widetilde A=0.387^{+0.020}_{-0.026}$ and $\alpha=0.036^{+0.034}_{-0.023}$. Thus the latter lies closer to the $\alpha=0$ background limit associated with $\Lambda$CDM. The correlation with $w_{{\rm gcg},0}$ reflects the requirement that the unified fluid remain matter-like at early epochs while producing late acceleration. Moreover, the four-chain comparison in Figure~\ref{fig:triang_plots} reveals clear supernova-likelihood dependence. The green PPS+gCg contour has the largest $H_0$ and the smallest $S_8$ central values among the four cases. Relative to PPS+$\Lambda$CDM, gCg raises $H_0$ by about $1.33\,{\rm km\,s^{-1}\,Mpc^{-1}}$ and lowers $S_8$ by about $0.011$. Both shifts occur in the directions needed to reduce the respective discrepancies, making this combination the most promising one in Figure~\ref{fig:triang_plots} for partial tension relief. Nevertheless, $H_0=68.64\pm0.52\,{\rm km\,s^{-1}\,Mpc^{-1}}$ remains well below the central SH0ES determination, while the $S_8$ posteriors substantially overlap. The result therefore indicates partial alleviation rather than a statistically established resolution. With DES--Dovekie, the smaller shifts show that this behaviour depends on the supernova likelihood. This sensitivity is consistent with the wider DESI-era lesson that evidence for extensions of $\Lambda$CDM can depend materially on data selection and calibration \cite{Li:2026asg}.

\begin{figure*}[t]
    \centering
    \begin{subfigure}[t]{0.48\linewidth}
        \centering
        \includegraphics[height=2.25in,width=\linewidth]{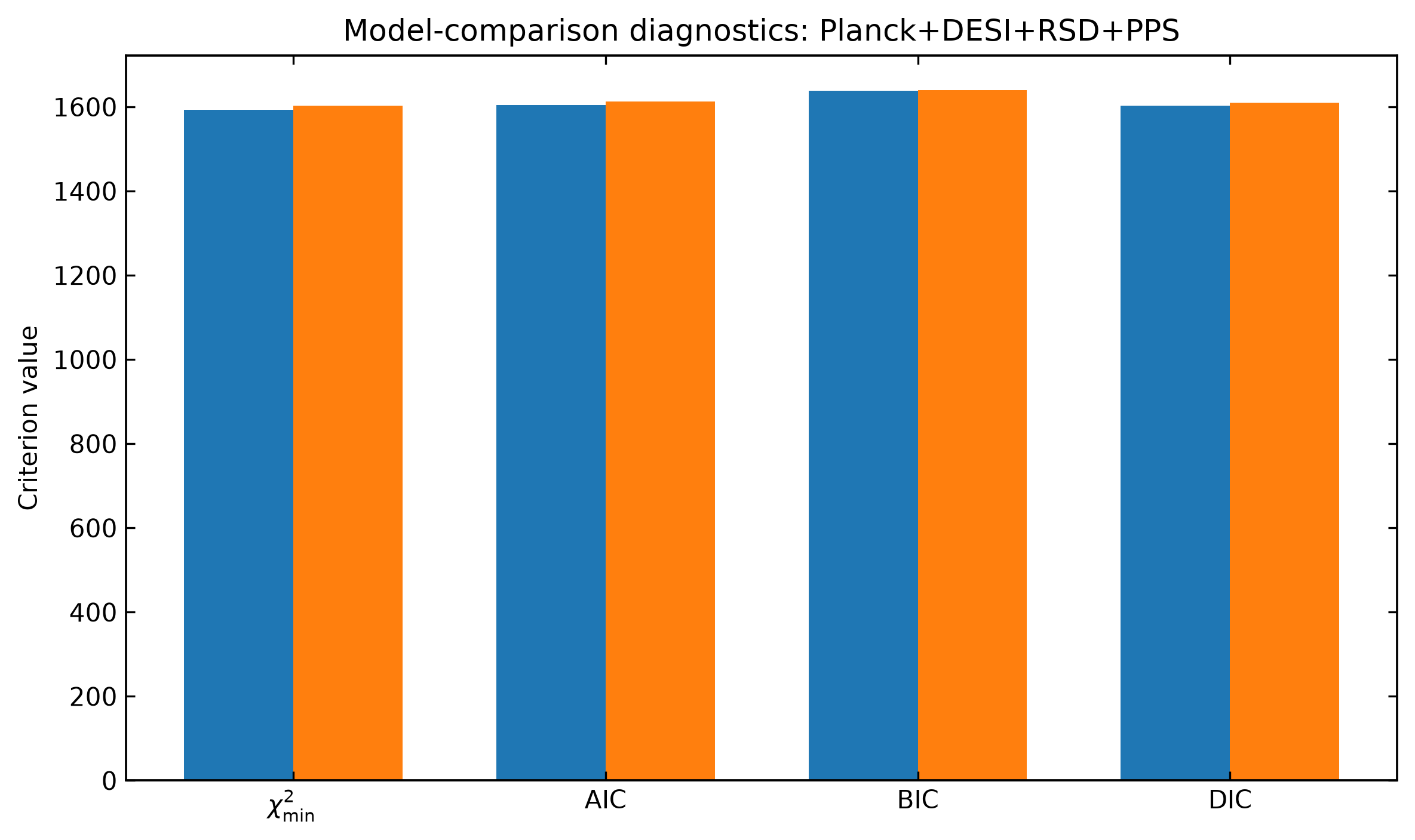}
        \caption{Planck+DESI+RSD+PPS.}
        \label{fig:model_pps}
    \end{subfigure}%
    \hfill
    \begin{subfigure}[t]{0.48\linewidth}
        \centering
        \includegraphics[height=2.25in,width=\linewidth]{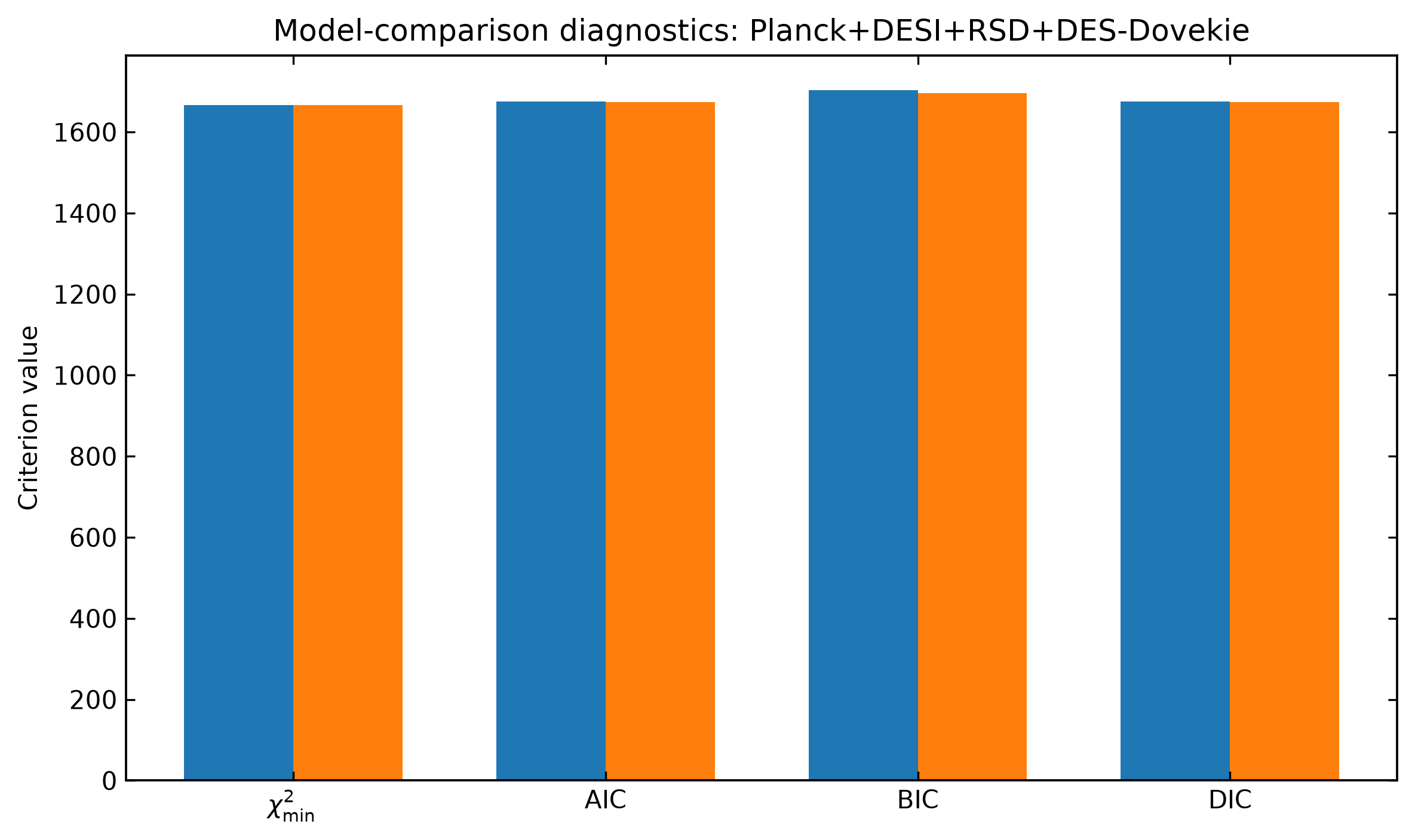}
        \caption{Planck+DESI+RSD+DES--Dovekie.}
        \label{fig:model_desd}
    \end{subfigure}
    \caption{Absolute fit and model-comparison diagnostics for entropic gCg and $\Lambda$CDM from the 800-live-point runs. Lower $\chi^2_{\min}$ indicates a better best fit, while lower AIC, BIC, and DIC values favour a model according to their respective complexity penalties.}
    \label{fig:tens4}
\end{figure*}

The model-comparison results should be read together with the CMB discussion in Sec.~\ref{subsec:planck_twocol}. The distance priors strongly constrain the background geometry but do not test the complete scale-dependent perturbation response of the unified fluid. Full Planck spectra would add independent constraints from the acoustic-peak structure, gravitational-potential evolution, polarization, and CMB lensing. They would probably tighten the constraints on $\alpha$ and $\widetilde A$ and could modify the favourable PPS+gCg displacement toward higher $H_0$ and lower $S_8$. Therefore, whether this simultaneous shift survives the full CMB spectral information remains an open numerical question rather than a consequence of the distance-prior analysis.

\begin{table*}[t]
\centering
\caption{Model-comparison diagnostics. The parameter count $k$ includes the profiled supernova magnitude/offset. The minimum $\chi^2$ is the lowest validated value from the posterior samples and dedicated best-fit search.}
\label{tab:model_comparison_raw}
\resizebox{\textwidth}{!}{%
\begin{tabular}{lccccccccc}
\hline\hline
Data set & Model & $k$ & $N_{\rm data}$ & $\ln Z$ & $\chi^2_{\min}$ & $\chi^2_{\nu}$ & AIC & BIC & DIC \\
\hline
Planck+DESI+RSD+PPS
& Entropic gCg
& 6 & 1724
& $-817.586 \pm 0.053$
& 1592.844
& 0.927
& 1604.844
& 1637.558
& 1603.138
\\
Planck+DESI+RSD+PPS
& $\Lambda$CDM
& 5 & 1724
& $-818.070 \pm 0.007$
& 1602.364
& 0.932
& 1612.364
& 1639.626
& 1610.250
\\
Planck+DESI+RSD+DES--Dovekie
& Entropic gCg
& 6 & 1843
& $-854.219 \pm 0.011$
& 1665.874
& 0.907
& 1677.874
& 1710.989
& 1674.987
\\
Planck+DESI+RSD+DES--Dovekie
& $\Lambda$CDM
& 5 & 1843
& $-850.008 \pm 0.007$
& 1666.348
& 0.907
& 1676.348
& 1703.944
& 1674.353
\\
\hline\hline
\end{tabular}%
}
\end{table*}

\begin{table*}[t]
\centering
\caption{Pairwise differences, defined as gCg minus $\Lambda$CDM. Negative information-criterion differences favour gCg, positive differences favour $\Lambda$CDM, and positive $\Delta\ln Z$ favours gCg. The Bayes factor is $B_{\rm gCg,\Lambda CDM}=\exp(\Delta\ln Z)$.}
\label{tab:model_comparison_delta}
\resizebox{\textwidth}{!}{%
\begin{tabular}{lcccccc}
\hline\hline
Data set & $\Delta\chi^2_{\min}$ & $\Delta$AIC & $\Delta$BIC & $\Delta$DIC & $\Delta\ln Z$ & $B_{\rm gCg,\Lambda CDM}$ \\
\hline
Planck+DESI+RSD+PPS
& $-9.520$
& $-7.520$
& $-2.068$
& $-7.112$
& $0.484\pm0.053$
& $1.62$
\\
Planck+DESI+RSD+DES--Dovekie
& $-0.474$
& $1.526$
& $7.045$
& $0.634$
& $-4.211\pm0.013$
& $1.48\times10^{-2}$
\\
\hline\hline
\end{tabular}%
}
\end{table*}

In this paper, we use very popular model selection criteria such as AIC \cite{akaike1974} that penalizes extra parameters according to
${\rm AIC}=\chi^2_{\rm min}+2k$. Here and throughout this discussion, the information-criterion differences are defined as
$\Delta{\rm IC}={\rm IC}_{\rm gCg}-{\rm IC}_{\Lambda{\rm CDM}}$,
so that negative values indicate a preference for the gCg model. Moreover, BIC \cite{schwarz1978} imposes a stronger penalty proportional to $\ln N_{\rm data}$. Similarly, DIC \cite{spiegelhalter2002} incorporates the effective posterior complexity through the Bayesian complexity parameter $p_D$. An interesting comparison of these model-selection methods and Bayesian evidence in cosmology is presented in \cite{RezaeiMalekjani2021}. The best-fit decomposition for PPS locates the improvement in which gCg changes the PPS, DESI, RSD, Planck-distance, and lensing contributions by approximately $-6.65$, $-7.11$, $+0.17$, $+4.00$, and $+0.08$, respectively. Thus the gain arises mainly from the low-redshift PPS and DESI terms and is partly opposed by the compressed Planck-distance term.

Tables~\ref{tab:model_comparison_raw} and \ref{tab:model_comparison_delta} separate absolute model scores from matched, within-dataset differences. For the sake of clarity, we point out that PPS has 1701 supernova entries, whereas DES--Dovekie has 1820. Adding 13 DESI BAO, 6 RSD, 3 Planck distance-prior, and 1 Planck lensing data points to the 1701 PPS or 1820 DES--Dovekie supernova entries gives $N_{\rm data}=1724$ and $1843$, respectively. In Table~\ref{tab:model_comparison_raw}, the PPS combination contains $N_{\rm data}=1724$ entries and gives $\chi^2_{\min}=1592.844$ for gCg versus $1602.364$ for $\Lambda$CDM and the corresponding reduced values are $0.927$ and $0.932$. The DES--Dovekie combination contains $1843$ entries and gives $1665.874$ versus $1666.348$, with both reduced values rounding to $0.907$. These reduced values show that neither best fit is visibly poor by this descriptive measure, whereas the substantial PPS improvement and the negligible DES--Dovekie improvement distinguish the two comparisons. Because the supernova samples and $N_{\rm data}$ differ, absolute $\chi^2$, information-criterion, and evidence values should be compared between models fitted to the \emph{same} data combination, not between PPS and DES--Dovekie rows.

In Table~\ref{tab:model_comparison_delta}, we show what survives the extra-parameter penalty. For PPS, $\Delta\chi^2_{\min}=-9.520$ more than offsets the AIC penalty of $2$ and the BIC penalty of $\simeq7.452$, yielding $\Delta\mathrm{AIC}=-7.520$ and $\Delta\mathrm{BIC}=-2.068$ and $\Delta\mathrm{DIC}=-7.112$ points in the same direction as AIC. The Bayes factor, however, is only $1.62$ ($\Delta\ln Z=0.484\pm0.053$) and its reported uncertainty is the numerical evidence error and does not turn this small ratio into strong support. For DES--Dovekie, the $0.474$ reduction in best-fit $\chi^2$ cannot compensate for the extra parameter with $\Delta\mathrm{AIC}=1.526$, $\Delta\mathrm{BIC}=7.045$, and $\Delta\mathrm{DIC}=0.634$. Thus the PPS likelihood rewards the extra gCg freedom at the best fit, but its integrated evidence remains inconclusive and DES--Dovekie yields little fit improvement and disfavors that added freedom. The evidence conclusions are conditional on the adopted prior ranges and compressed-CMB likelihood and require a prior-sensitivity check before a broader model-selection claim.

Figure~\ref{fig:tens2} shows how the inferred parameters change
when entropic gCg replaces $\Lambda$CDM within each data
combination. For PPS, the posterior-difference distribution is
displaced toward higher $H_0$, lower $\Omega_m$, and lower $S_8$,
with a center near
$(\Delta H_0,\allowbreak\Delta\Omega_m,\allowbreak\Delta S_8)
=\allowbreak(1.33,\allowbreak-0.0083,\allowbreak-0.011)$.
The DES--Dovekie distribution shows the same directions of
change but smaller shifts, near
$(0.44,\allowbreak-0.0031,\allowbreak-0.0046)$.
Thus, the parameter displacement favourable to simultaneous
$H_0$ and $S_8$ alleviation depends appreciably on the supernova
likelihood. These distributions compare models fitted to the
same data but they do not measure the statistical significance
of either cosmological tension. Finally,
Figure~\ref{fig:tens4} visualizes the absolute $\chi^2_{\min}$,
AIC, BIC, and DIC columns of
Table~\ref{tab:model_comparison_raw}, complementing the
within-dataset differences reported in
Table~\ref{tab:model_comparison_delta}, and makes change in the
model ranking visible across the two supernova datasets.
In the PPS panel, gCg has a substantially lower best-fit
$\chi^2_{\min}$ than $\Lambda$CDM ($1592.844$ versus $1602.364$).
That improvement is large enough that gCg also has lower AIC,
BIC, and DIC values, although its BIC advantage is modest.
In the DES--Dovekie panel, the best-fit values are almost
identical ($1665.874$ versus $1666.348$), and the extra gCg
parameter then leaves $\Lambda$CDM with lower AIC, BIC, and DIC
values. Since the DIC difference there is small, so it should
not be described as a strong preference.

In addition, the posterior shifts and model-comparison results identify a data-dependent opportunity for entropic gCg rather than a robust preference for it. With PPS, the model moves $H_0$ and $S_8$ in favourable directions and improves the best fit enough to overcome the information-criterion penalties, yet the Bayes factor remains close to unity. With DES--Dovekie, the shifts and fit improvement are smaller, and the Bayesian evidence favours $\Lambda$CDM under the adopted priors. The present analysis therefore establishes which likelihood choice drives the apparent advantage and where it is lost. Whether the PPS-favoured parameter region remains supported under different priors and a full CMB treatment of the entropic perturbations warrants further investigation.

\section{Concluding remarks}

In this work, we investigated an entropic UDM generalized Chaplygin-gas cosmology through a combined analysis of background evolution, linear growth, and observational tension diagnostics. The model was confronted with PPS and DES--Dovekie supernova compilations, DESI BAO measurements, RSD data, compressed Planck 2018 distance priors, and a Planck lensing prior.  Our main goal was to determine whether a unified dark fluid can reshape the parameter degeneracies associated with current cosmological discrepancies while remaining compatible with the combined datasets.

At the background level, we have shown that the UDM gCg model reproduces a cosmic expansion history close to the standard $\Lambda$CDM scenario while introducing controlled late-time deformations in the distance-redshift relation. The evolution of the Hubble parameter $H(z)$, the deceleration parameter $q(z)$, and the effective EoS $w(z)$ demonstrate that the model naturally interpolates between a matter-dominated regime at high redshift and an accelerated DE-like phase at late times. In particular, the transition of EoS from $w \simeq 0$ to $w \rightarrow -1$ reinforces the unified-dark-sector interpretation of the model, showing that a single fluid can effectively mimic both CDM and DE during different cosmological epochs.

An important result is that the model comparison depends on the supernova compilation. With PPS, gCg improves the best fit and is preferred by AIC and DIC, while BIC gives only a modest advantage and the Bayesian evidence remains inconclusive. With DES--Dovekie, the best-fit improvement is negligible, i.e., BIC and Bayesian evidence favour $\Lambda$CDM, while AIC and DIC show only small differences. We point out that the reported evidence errors quantify numerical integration uncertainty, not sensitivity to priors or likelihood choice. The analysis therefore does not support a dataset-independent statistical preference for entropic gCg. However, it identifies the late-time supernova likelihood as an important factor controlling the comparison.

At the perturbative level, we have shown that the adiabatic gCg model leads to a non-negligible sound speed that can suppress structure formation and generate oscillatory behaviour in the matter-overdensity evolution. This motivates the entropic prescription adopted in the present work, where the effective rest-frame sound speed satisfies $c_{\rm eff}^2\simeq0$, allowing the unified dark fluid to cluster approximately as CDM. Although this prescription improves the phenomenological viability of the model, perturbative observables remain more restrictive than the background sector and require further refinement for a fully consistent cosmology.

To compare the posterior distributions obtained from the two cosmological models, we used \texttt{Tensiometer} \cite{Raveri2020Tensiometer}. These calculations quantify cross-model posterior displacement under the same joint likelihood, rather than an early-versus-late tension probability. The PPS+gCg solution is nevertheless noteworthy because it moves $H_0$ upward and $S_8$ downward at the same time. Among the four cases displayed in Fig.~\ref{fig:triang_plots}, it is the configuration most favourable for simultaneously alleviating both discrepancies. The shifts are moderate, however, and a quantitative tension significance would require separate early- and late-Universe posteriors within each cosmology.

Our results suggest that entropic gCg remains a useful theoretical laboratory for unified-dark-sector cosmology. It generates moderate late-time shifts in the directions required for joint $H_0$ and $S_8$ alleviation and gives a better PPS maximum likelihood, yet its evidence advantage is inconclusive and is not reproduced with DES--Dovekie. A robust assessment requires controlled prior tests and a full treatment of cosmological perturbations. The background EoS $w(a)$ does not uniquely determine the perturbation dynamics of an entropic Chaplygin fluid and a complete implementation requires evolution of the density and velocity perturbations, a covariant non-adiabatic pressure prescription, consistent initial conditions, and stable treatment of the $w\rightarrow-1$ regime. Once this sector is incorporated into an Einstein--Boltzmann solver, the full Planck TT, TE, EE, and lensing likelihoods will determine whether the parameter region producing the simultaneous upward shift in $H_0$ and downward shift in $S_8$ remains compatible with the acoustic-peak structure and CMB-lensing spectrum. Depending on the perturbation dynamics, the full likelihood may either tighten the present background-level indication or constrain the additional gCg freedom more strongly. A subsequent work on this matter is on progress and will be reported elsewhere.

\begin{acknowledgements}

K. A. K. thanks Capes-Brazil for the financial support. CHC-A and AJSC are very thankful for the financial support of the NAPI ``Fenômenos Extremos do Universo'' of Fundação de Apoio à Ciência, Tecnologia e Inovação do Paraná, (NAPI FÍSICA – FASE 2), under protocol No 22.687.035-0. CHC-A also acknowledges the financial support of Fundação de Apoio à Ciência, Tecnologia e Inovação do Paraná (Grant No. PRD2023361000304). AJSC acknowledges Conselho Nacional de Desenvolvimento Cient\'{i}fico e Tecnologico (CNPq) for the partial financial support for this work (Grant No. 305881/2022-1) and Fundação da Universidade Federal do Paraná (FUNPAR, Paraná Federal University Foundation) through public notice 04/2023 - Pesquisa / PRPPG / UFPR for the partial financial support (Process No. 23075.019406/2023-92). 

This article is based upon work from the COST Action CA21136 ``Ad
dressing observational tensions in cosmology with systematics and fundamental physics'' (CosmoVerse), supported by COST (European Cooperation in Science and Technology).

\end{acknowledgements}

\bibliographystyle{spphys}       
\bibliography{sn-bibliography}   

%

\end{document}